\documentclass[twocolumn]{aastex61}
\pdfoutput=1 
\usepackage{amsmath,amstext}
\usepackage[T1]{fontenc}
\usepackage[figure,figure*]{hypcap}
\usepackage{longtable}

\shorttitle{AASTeX 6.1 Template}
\shortauthors{Rocha \& Pilling }

\begin{document}

\title{Tracking the evolutionary stage of protostars by the abundances of astrophysical ices}

\author{W. R. M. Rocha}
\affiliation{Instituto de Pesquisa \& Desenvolvimento (IP\&D), Universidade do Vale do Para{\'{i}}ba, Av. Shishima Hifumi 2911, 12244-000, S{\~{a}}o Jos{\'{e}} dos Campos, SP, Brazil}
\affiliation{Niels Bohr Institute \& Centre for Star and Planet Formation, University of Copenhagen, {\O}ster Voldgade 5-7, DK-1350 Copenhagen K., Denmark}
\author{S. Pilling}
\affiliation{Instituto de Pesquisa e Desenvolvimento (IP\&D), Universidade do Vale do Para{\'{i}}ba, Av. Shishima Hifumi 2911, 12244-000, S{\~{a}}o Jos{\'{e}} dos Campos, SP, Brazil}
\affiliation{Departamento de F{\'{i}}sica, Instituto Tecnol{\'{o}}gico de Aeronautica, ITA - DCTA, Vila das Ac{\'{a}}cias, S{\~{a}}o Jos{\'{e}} dos Campos, 12228-900, SP-Brazil}

\begin{abstract}
The physical evolution of Young Stellar Objects (YSOs) is accompanied by an enrichment of the molecular complexity, mainly triggered by the heating and energetic processing of the astrophysical ices. In this paper, a study of how the ice column density varies across the protostellar evolution has been performed. Tabulated data of H$_2$O, CO$_2$, CH$_3$OH, HCOOH observed by ground- and space-based telescopes toward 27 early-stage YSOs were taken from the literature. The observational data shows that ice column density and spectral index ($\alpha$), used to classify the evolutionary stage, are well correlated. A 2D continuum radiative transfer simulation containing bare and grains covered by ices at different levels of cosmic-ray processing were used to calculate the Spectral Energy Distributions (SEDs) in different angle inclinations between face-on and edge-on configuration. The H$_2$O:CO$_2$ ice mixture was used to address the H$_2$O and CO$_2$ column density variation whereas the CH$_3$OH and HCOOH are a byproduct of the virgin ice after the energetic processing. The simulated spectra were used to calculate the ice column densities of YSOs in an evolutionary sequence. As a result, the models show that the ice column density variation of HCOOH with $\alpha$ can be justified by the envelope dissipation and ice energetic processing. On the other hand,  the ice column densities are mostly overestimated in the cases of H$_2$O, CO$_2$ and CH$_3$OH, even though the physical and cosmic-ray processing effects are taken into account. 
\end{abstract}

\keywords{Astrochemistry (75), Infrared sources (793), Interstellar abundances (832), Laboratory astrophysics (2004), Star forming regions (1565)}

\section{Introduction}
Star formation begins in very embedded regions of Molecular Clouds with no measurable flux in the visible light. Nevertheless, the dust heated by the internal Young Stellar Object (YSOs) emits radiation in infrared (IR) wavelengths. An IR-based classification of protostars have been proposed by \citet{Lada1984, Lada1987, Andre1993, Greene1994} and takes into account the IR excess in the Spectral Energy Distributions (SEDs). Formally, the method calculates the spectral index between 2~$\mu$m and 24~$\mu$m given by:
\begin{equation}
\alpha \equiv \dfrac{d\mathrm{log\lambda F{_\lambda}}}{d \mathrm{log\lambda}}
\end{equation}
where $\lambda$ is the wavelength and F$_\lambda$ the flux. The physical structure of each evolutionary stage was reviewed by \citet{Williams_Cieza2011}, and the relation with $\alpha$ is also discussed: (i) Class I objects are characterized by a large spherical envelope and a disk, with no optical or near-IR emission and $\mathrm{\alpha_{Class \; I} > 0.3}$; (ii) Class II are T-Tauri objects with an optically thick disk in the visible, but strong UV emission at short-wavelengths and $\mathrm{-0.3 > \alpha_{Class \; II} > -1.6}$; (iii) The last stage is defined by Class III objects characterized by a very low infrared excess and $\mathrm{\alpha_{Class \; III} < -1.6}$. Since Class 0 YSOs are extremely embedded objects, they are not classified by the spectral index. Its classification is, however, given by the ratio between the sub-millimeter luminosity ($L_{\mathrm{smm}}$) and the bolometric luminosity ($L_{\mathrm{bol}}$) as proposed by \citet{Andre1993}.

In addition to the physical evolution of YSOs, astrophysical ices play an important role in enriching the chemical complexity in the Interstellar medium. As shown in \citet{Boogert2015}, given the conditions of large H/CO gas ratio, T~$>$~20~K, n~$\geq$~10$^3$~cm$^{-3}$, $A_{\mathrm{V}}$~$>$~1.5, the hydrogenation mechanism leads to the saturation of adsorbed atoms on the dust grain, and the H$_2$O, CH$_4$, NH$_3$ ices are formed. On the other hand, if H/CO gas ratio is small, T $<$ 20 K, $n$~$\geq$~10$^4$~cm$^{-3}$ and $A_{\mathrm{V}}$~$>$~3 mag, the CO accretion rate is increased. Thereafter, surface reactions between CO and oxygen atoms might lead to the formation of CO$_2$ ice. In low-density regions, however, CO$_2$ ice can be formed via CO~+~OH reactions \citep{Allamandola1988}. In regions denser than 10$^4$~cm$^{-3}$, the catastrophic CO freeze-out leads to the efficient formation of methanol ice at low temperatures via carbon monoxide hydrogenation \citep{Watanabe2003, Fuchs2009}.

Besides the adsorption mechanisms, the chemical enrichment takes place by thermal and energetic processing of the icy compounds. For instance, at around 30~K the dust grains are warm enough to enable diffusion and recombination of small molecules on the ice, and complex organic molecules (COMs) are formed \citep{Garrod_Herbst2006, Herbst_vanDishoeck2009, Caselli_Ceccarelli2012}. On the other hand, the energetic processing of ices is the main route to form complex species as reviewed by \citet{Oberg2016} for photochemical induced processes and particles by \citet{Boduch2015, Rothard2017}.

In order to investigate how the abundance of ice species are related to the evolutionary sequence, tabulated data of ice column density and spectral index of 27 YSOs were taken from \citet{Pontoppidan2008, Boogert2008}. Additionally, 2D continuum radiative transfer models of protostars in an evolutionary sequence, containing dust grains covered by ice at different levels of processing was used to simulate how the ice column density varies with the protostar evolution.  

The paper is structured as follows: Section 2 shows the source sample addressed in this paper, and Section~3 shows the ice column density decreasing with the spectral index i.e. as the protostar evolves. Section~4 characterizes the radiative transfer simulations and the ice-dust model and shows how the laboratory data were employed in the simulations. In Section~5 the results and discussion are shown, and the conclusions are in Section~6. Appendix A detail how the IR range between 5.5$-$7.5~$\mu$m was decomposed using Gaussian profiles.

\section{Source sample}
Table \ref{tab:dataset} lists 27 well-known Low-Mass Young Stellar Objects (LYSOs) addressed in this paper. Both the evolutionary stage and the column densities for these sources were directly taken from \citet{Boogert2008} and \citet{Pontoppidan2008}. The authors used broadband (2.17$-$24~$\mu$m) classification scheme to calculate the spectral index through Equation 1, and cover Class 0/I to Class I/II. The spectral index error bar was not provided.

Spitzer data in the mid-IR, combined with ground-based observations in the near-IR (Keck NIRSPEC \citep{McLean1998} and VLT-ISAAC \citep{Moorwood1997}), when available, were used to calculate the ice column densities the sources listed in Table \ref{tab:dataset}. The IR spectrum was converted into optical depth using the equation $\tau_{\lambda} = -\mathrm{ln}\left(\frac{F_{\lambda}^{obs}}{F_{\lambda}^{cont}}\right)$ to allow the column density calculation, where $F_{\lambda}^{obs}$ is the observed flux and $F_{\lambda}^{cont}$ is the SED continuum. The continuum was defined by a low-order polynomial for all the sources. Thereafter, the ice column density was calculated by:
\begin{equation}
    N_{ice} = \frac{1}{\mathcal{A}} \int_{\nu_1}^{\nu_2} \tau_{\nu} d\nu
    \label{CD}
\end{equation}
where $\mathcal{A}$ is the band strength of a specific vibrational mode and $\nu$ is the wavenumber in units of cm$^{-1}$. 

In \citet{Boogert2008}, the H$_2$O ice column density was calculated from the O$-$H stretching mode at 3.0~$\mu$m, or the libration mode at 12.3~$\mu$m, after the removal of the contribution of silicate absorption. The bending mode at 6.0~$\mu$m has been avoided, since such a band cannot be attributed only to water as reported in \citet{Gibb2000, Keane2001, Gibb2004}. Methanol ice column density was derived from the absorption features at 3.54~$\mu$m and 9.7~$\mu$m. Formic acid (HCOOH), has many vibrational modes in the IR, but its column density was calculated from the band at 7.25~$\mu$m, since the other modes are blended with H$_2$O and other alcohols such as ethanol and methanol. The carbon dioxide ice column density in \citet{Pontoppidan2008}, was calculated from the bending mode at 15.2~$\mu$m. Table \ref{tab:columndens} shows the column densities calculated for the ices mentioned above.

\startlongtable
\begin{deluxetable*}{c c c c c c c}
\tablecaption{Source sample\label{tab:dataset}}
\tablehead{\colhead{Source} & \colhead{R.A. (J2000)} & \colhead{DEC (J2000)} & \colhead{Cloud} & \colhead{$\alpha_{IR}$} & \colhead{Stage} & \colhead{Telescope}\\
}
\centering
\startdata
IRAS 03245+3002{\dotfill}  & 03h27'39.03'' & +30$^{\circ}$12'59.3'' & Perseus   & 2.70  & Class 0/I & Spitzer \\
L1455 SMM 1{\dotfill}  & 03h27'43.25'' & +30$^{\circ}$12'28.8'' & Perseus   & 2.41  & Class 0/I & Spitzer \\
IRAS 03271+3013{\dotfill}  & 03h30'15.16'' & +30$^{\circ}$23'48.8'' & Perseus   & 2.06  & Class 0/I & Spitzer, Keck \\
B1-c{\dotfill}  & 03h33'17.89'' & +31$^{\circ}$09'31.0'' & Perseus   & 2.66  & Class 0/I & Spitzer \\
HH 46 IRS{\dotfill}  & 08h25'43.78'' & -51$^{\circ}$00'35.6'' & HH 46   & 1.70  & Class 0/I & Spitzer, VLT \\
CRBR 2422.8-3423{\dotfill}  & 16h27'24.61'' & -24$^{\circ}$41'03.3'' & Oph   & 1.60  & Class 0/I & Spitzer, Keck \\
SSTc2d J171122.2-272602{\dotfill}  & 17h11'22.16'' & -27$^{\circ}$26'02.3'' & B59   & 2.26  & Class 0/I & Spitzer \\
2MASS J17112317-2724315{\dotfill}  & 17h11'23.13'' & -27$^{\circ}$24'32.6'' & B59   & 2.48  & Class 0/I & Spitzer, Keck \\
CrA IRS 7A{\dotfill}  & 19h01'55.32'' & -36$^{\circ}$57'22.0'' & CrA   & 2.23  & Class 0/I & Spitzer, VLT \\
CrA IRAS 32{\dotfill}  & 19h02'58.69'' & -37$^{\circ}$07'34.5'' & CrA   & 2.15  & Class 0/I & Spitzer \\
L1014 IRS{\dotfill}  & 21h24'07.51'' & +49$^{\circ}$59'09.9'' & L1014   & 1.60  & Class 0/I & Spitzer, Keck \\
L1489 IRS{\dotfill} & 04h04'43.07'' & +26$^{\circ}$18'56.4'' & Taurus   & 1.10  & Class I & Spitzer, Keck \\
HH 300{\dotfill} & 04h26'56.30'' & +24$^{\circ}$43'35.3'' & Taurus   & 0.79  & Class I & Spitzer, Keck \\
DG Tau B{\dotfill} & 04h27'02.66'' & +26$^{\circ}$05'30.5'' & Taurus   & 1.16  & Class I & Spitzer, Keck \\
IRAS 12553-7651{\dotfill} & 12h59'06.63'' & -77$^{\circ}$07'40.0'' & Cha   & 0.76  & Class I & Spitzer \\
Elias 29{\dotfill} & 16h27'09.42'' & -24$^{\circ}$37'21.1'' & Oph   & 0.53  & Class I & ISO \\
IRAS 17081-2721{\dotfill} & 17h11'17.28'' & -27$^{\circ}$25'08.2'' & B59   & 0.55  & Class I & Spitzer, Keck \\
EC 82{\dotfill} & 18h29'56.89'' & +01$^{\circ}$14'46.5'' & Serpens   & 0.38  & Class I & Spitzer\\
SVS 4-5{\dotfill} & 18h29'57.59'' & +01$^{\circ}$13'00.6'' & Serpens   & 1.26  & Class I & Spitzer, VLT \\
R CrA IRS 5{\dotfill} & 19h01'48.03'' & -36$^{\circ}$57'21.6'' & CrA   & 0.98  & Class I & Spitzer, VLT \\
HH 100 IRS{\dotfill} & 19h01'50.56'' & -36$^{\circ}$58'08.9'' & CrA   & 0.80  & Class I & ISO \\
RNO 15{\dotfill} & 03h27'47.68'' & +30$^{\circ}$12'04.3'' & Perseus   & -0.21  & Class I/II & Spitzer, Keck \\
IRAS 13546-3941{\dotfill} & 13h57'38.94'' & -39$^{\circ}$56'00.2'' & BHR 92   & -0.06  & Class I/II & Spitzer \\
RNO 91{\dotfill} & 16h34'29.32'' & -15$^{\circ}$47'01.4'' & L43   & 0.03  & Class I/II & Spitzer, VLT \\
EC 74{\dotfill} & 18h29'55.72'' & +01$^{\circ}$14'31.6'' & Serpens   & -0.25  & Class I/II & Spitzer, Keck \\
EC 90{\dotfill} & 18h29'57.75'' & +01$^{\circ}$14'05.9'' & Serpens   & -0.09  & Class I/II & Spitzer \\
CK 4{\dotfill} & 18h29'58.21'' & +01$^{\circ}$15'21.7'' & Serpens   & -0.25  & Class I/II & Spitzer \\
\enddata
\end{deluxetable*}

\startlongtable
\begin{deluxetable*}{c c c c c}
\tablecaption{Ice column density of the Source sample.
\label{tab:columndens}}
\tablehead{\colhead{Source} & \colhead{$\mathrm{N_{H_2O}}$~$^a$} & \colhead{$\mathrm{N_{CO_2}}$~$^c$} & \colhead{$\mathrm{N_{CH_3OH}}$~$^{a}$} & \colhead{$\mathrm{N_{HCOOH}}$~$^{a,e}$}\\
 & (10$^{18}$ cm$^{-2}$) & (10$^{18}$ cm$^{-2}$)& (10$^{18}$ cm$^{-2}$)& (10$^{18}$ cm$^{-2}$)
}
\centering
\startdata
IRAS 03245+3002{\dotfill}  & 39.31$\pm$5.65$^b$ & - & 3.85 & 0.47 \\
L1455 SMM 1{\dotfill}  & 18.21$\pm$2.82$^b$ & 6.34$\pm$0.44 & 2.45  & 0.60$\pm$0.02\\
IRAS 03271+3013{\dotfill}  & 7.69$\pm$1.76$^b$ & 1.53$\pm$0.09 & 0.43 & 0.19\\
B1-c{\dotfill}  & 29.55$\pm$5.65$^b$ & 8.4 & 2.1 & 0.35$\pm$0.01\\
HH 46 IRS{\dotfill}  & 7.79$\pm$0.77 & 2.16$\pm$0.01 & 0.42$\pm$0.01$^d$ & 0.21$\pm$0.01\\
CRBR 2422.8-3423{\dotfill}  & 4.19$\pm$0.41 & 1.05$\pm$0.01 & 0.38 & - \\
SSTc2d J171122.2-272602 {\dotfill} & 13.94$\pm$2.92$^b$ & - & 0.18 & 0.41$\pm$0.03  \\
2MASS J17112317-2724315{\dotfill}  & 19.49$\pm$0.23$^b$ & - & 0.62  & 0.48$\pm$0.16 \\
CrA IRS 7A{\dotfill}  & 10.89$\pm$1.92$^b$ & 1.96$\pm$0.12 & 0.41  & -\\
CrA IRAS 32{\dotfill}  & 5.26$\pm$1.88$^b$ & 1.87$\pm$0.21 & 0.95 & -\\
L1014 IRS{\dotfill}  & 7.16$\pm$0.91$^b$ & - & 0.22$\pm$0.05  & - \\
L1489 IRS{\dotfill} & 4.26$\pm$0.51 & 1.62$\pm$0.02 & 0.21$\pm$0.01  & 0.12\\
HH 300{\dotfill} & 2.59$\pm$0.25 & - & 0.17 & 0.06\\
DG Tau B{\dotfill} & 2.29$\pm$0.39 & 0.54 & 0.13   & 0.07\\
IRAS 12553-7651{\dotfill} & 2.98$\pm$0.56$^b$ & 0.61$\pm$0.01 & 0.08  & 0.05 \\
Elias 29{\dotfill} & 3.04$\pm$0.30& 0.84$\pm$0.06 & 0.14$^d$ & 0.04\\
IRAS 17081-2721{\dotfill} & 1.31$\pm$0.13 & - & 0.04$^d$ & 0.03\\
EC 82{\dotfill} & 0.39$\pm$0.07 & 0.25$\pm$0.01 & 0.05  & 0.01\\
SVS 4-5{\dotfill} & 5.65$\pm$1.13 & 1.72$\pm$0.05 & 1.41$\pm$0.19$^d$ & - \\
R CrA IRS 5{\dotfill} & 3.58$\pm$0.26 & 1.42$\pm$0.02 & 0.23$\pm$0.04 & 0.15\\
HH 100 IRS{\dotfill} & 2.45$\pm$0.24 & - & 0.23$^d$ & 0.06\\
RNO 15{\dotfill} & 0.69$\pm$0.06 & 0.25$\pm$0.01 & 0.03$^d$  & 0.04\\
IRAS 13546-3941{\dotfill} & 2.07$\pm$0.21$^b$ & 0.87$\pm$0.02 & 0.08  & -\\
RNO 91{\dotfill} & 4.25$\pm$0.36 & 1.16$\pm$0.02 & 0.24  & -\\
EC 74{\dotfill} & 1.07$\pm$0.18 & 0.30$\pm$0.05 & 0.1$^d$  & 0.03\\
EC 90{\dotfill} & 1.69$\pm$0.16 & 0.54$\pm$0.05 & 0.11$\pm$0.01 & 0.06\\
CK 4{\dotfill} & 1.50$\pm$0.01 & 0.20$\pm$0.01 & - & -\\
\enddata
\tablecomments{$^a$ Taken from \citet{Boogert2008}. \\
$^b$ $N_{\mathrm{H_2O}}$ calculated from the H$_2$O libration mode at 13.6~$\mu$m using the band strength $\mathcal{A_{\mathrm{H_2O}}}$~=~2.8 $\times$~10$^{-17}$ cm molecule$^{-1}$. The stretching mode at 3~$\mu$m was used in the other cases.\\
$^c$ Taken from \citet{Pontoppidan2008}. \\
$^d$ The column densities was calculated from the absorption features at 3.53~$\mu$m, using the band strength of $\mathcal{A_{\mathrm{CH_3OH}}}$~=~5.6 $\times$~10$^{-18}$~cm~molecule$^{-1}$. The absorption at 9.7~$\mu$m was used in the other cases, assuming the band strength of $\mathcal{A_{\mathrm{CH_3OH}}}$~=~1.6 $\times$~10$^{-17}$~cm~molecule$^{-1}$.\\
$^e$ The column densities was calculated from the absorption features at 7.25~$\mu$m, using the band strength of $\mathcal{A_{\mathrm{HCOOH}}}$~=~1.5 $\times$~10$^{-17}$~cm~molecule$^{-1}$.}
\end{deluxetable*}

\section{Correlations between ice column density and spectral index}
\begin{figure*}
\includegraphics[width=\textwidth,height=\textheight,keepaspectratio]{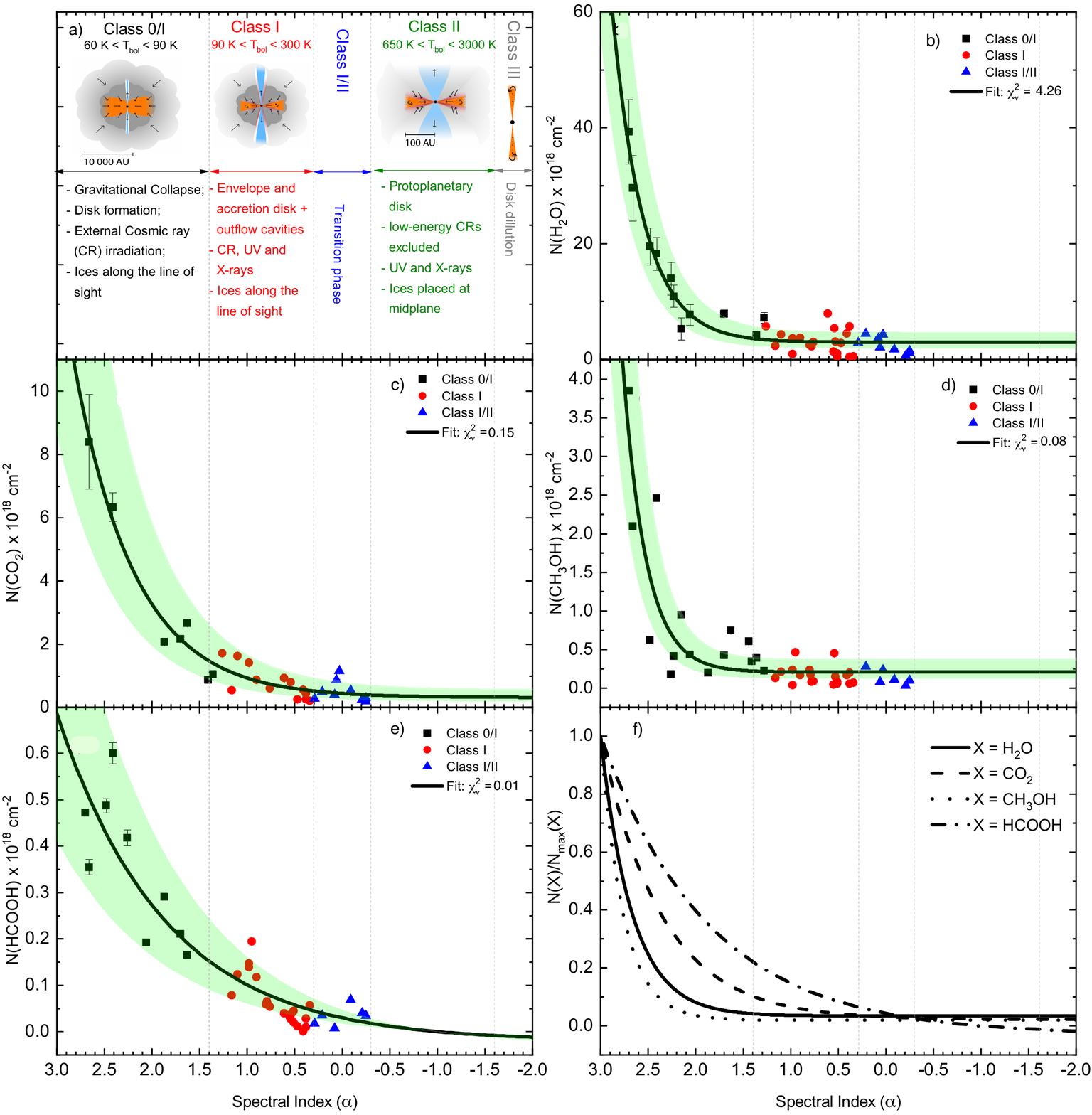}
\caption{Column density correlation of the frozen molecules H$_2$O, CO$_2$, CH$_3$OH, HCOOH with the spectral index ($\alpha$). Panel {\it a} shows a schematic representation of the structure of YSO for each evolutionary stage \citep{Persson2014}. The colours and symbols in panels ({\it b} - {\it e}) represent the evolutionary stages of protostars and the green shaded area corresponds to 99\% confidence interval. The exponential fit is shown by the solid black line. Panel {\it f} compares all fits for the normalized data.}
\label{fig:abs_alpha}
\end{figure*}

Figure \ref{fig:abs_alpha} shows the ice column density and the spectral index ($\alpha$) for the Class 0/I, Class I and Class I/II, given by the black squares, red circles and blue triangles, respectively. H$_2$O is the most abundant ice toward all the sources, followed by CO$_2$, CH$_3$OH and HCOOH, which agrees with the expected abundances shown by \citet{Oberg2011}.

Very early stages of the protostellar evolution are dominated by a cold envelope under gravitational collapse. \citet{Jorgensen2009} found evidence of envelope dissipation, estimated from the increasing of the disk-envelope mass ratio in ten Class 0 and Class I systems. Such a trend has also been observed in \citet{Andersen2019} toward the Perseus Molecular Cloud, where they also found strong evidence of the disk growth at Class 0 stage.

Since the ices are formed onto dust grains in cold regions, the dusty envelope dissipation might lead to an ice column density decreasing, as suggested in Figure \ref{fig:abs_alpha}. In order to make easier the comparison between $N_{\mathrm{ice}}$ and $\alpha$ in this paper, an exponential function given by the Equation~3 has been assumed:
\begin{equation}
    N_{\mathrm{ice}} = N_{\mathrm{ice}}^{plateau} + s \cdot e^{r \cdot \alpha} 
\end{equation}
where $N_{\mathrm{ice}}^{plateau}$ is the ice column density {\it plateau} toward the line of sight, $s$ is a scale factor of the amplitude variation in the y-axis and $r$ the decreasing rate (positive since x-axis is inverted). Since the presence of foreground clouds toward star-forming regions has been reported in the literature \citep{Boogert2002, Pontoppidan2005, vanDishoeck2011, Smith2015}, the term $N_{\mathrm{ice}}^{plateau}$ aims to take into account this factor. To compare the goodness of the fit among the ice species, the reduced $\chi^2$ was used, given by $\chi_{\nu}^2 = \chi^2/(n-m)$, where $n$ is the number of data points, and $m$ is the degree of freedom. Acceptable fits requires $\chi_{\nu}^2 \sim \mathrm{1}$. 

Table~\ref{tab:param} shows the parameters obtained from the exponential fit as well as the $\chi_{\nu}^2$. The $N_{\mathrm{H_2O}}^{plateau}$ and $N_{\mathrm{CO_2}}^{plateau}$ are in agreement with the ice column densities in foreground clouds proposed in the literature. In fact, \citet{Boogert2000} suggest that only 30\% of the ice column density observed toward Elias~29 protostar belong to the object itself, whereas a {\it plateau} of around $N_{\mathrm{H_2O}} \sim 2 \times 10^{18}$ cm$^{-2}$ is hosted by foreground clouds. Toward CRBR~2422.8$-$3423, \citet{Pontoppidan2005} estimate that, at least, 50\% of the H$_2$O ice column density lies in foreground clouds, namely, $N_{\mathrm{H_2O}} \leq 1.8 \times 10^{18}$ cm$^{-2}$. For face-on Class II YSO 2MASSJ 1628137 in Taurus, the estimated ice column density is about one order of magnitude lower than for the previous Class I objects, i.e. $N_{\mathrm{H_2O}} \sim 0.2 \times 10^{18}$ cm$^{-2}$ \citep{Aikawa2012}. In the case of CO$_2$ ice, the estimated column density in foreground clouds are $N_{\mathrm{CO_2}} \sim 2 \times 10^{17}$ cm$^{-2}$ in Elias~29 \citep{Boogert2000} and $N_{\mathrm{CO_2}} \leq 6 \times 10^{17}$ cm$^{-2}$ \citep{Pontoppidan2005} in CRBR~2422.8$-$3423. In the case of  CH$_3$OH and HCOOH, the derived parameters must be used with caution due to the poor fit indicated by the $\chi_{\nu}^2$.

From the decreasing rate ($r$), one can observe that CO$_2$ ice decreases slower than H$_2$O with the protostellar evolution. Due to the lower thermal desorption of pure CO$_2$ ($\sim$75~K), the most probable scenario in the Interstellar Medium is that C-rich molecules are trapped in a H$_2$O ice matrix. In this case, part of the CO$_2$ ice is only released back to the gas-phase at around 150~K, that characterizes the H$_2$O temperature desorption \citep{Collings2004}. In addition, the CO$_2$ ice can be formed via a surface reaction between OH and CO as shown by \citet{Allamandola1988}. As a consequence, in an environment dominated by radiation, CO$_2$ is slowly destroyed compared to H$_2$O because of its reformation mechanism from the water photo-products. Regarding formic acid and methanol, although the fits indicate that their respective ice column densities drop slower and faster with the spectral index, this result can be called into question due to the poor exponential fit. If, however, this is a real effect, a likely explanation is the efficient HCOOH formation both in gas- and solid-phase compared to CH$_3$OH \citet{Agundez2013}. To allow easier comparison between all the fits, Figure~\ref{fig:abs_alpha}f shows the normalized ice column densities for the four ices addressed in this paper.

In order to address if the envelope dissipation, viewing angle and the energetic processing of ices are able to provide a likely explanation for the ice column density variation shown in Figure~\ref{fig:abs_alpha}, a computational model of YSOs at different stages surrounded by dust and ice was employed in this paper. To mimic the energetic processing of the ices across the protostellar evolution, laboratory data of H$_2$O:CO$_2$ taken from \citet{Pilling2010} and \citet{Rocha2017} has also been used.

\begin{table*}
\caption{Parameters obtained from the exponential fit in Figure \ref{fig:abs_alpha}}
\centering
\label{tab:param}
\begin{tabular}{c c c c c}
\hline\hline
Ice specie & $\mathrm{N_{\mathrm{ice}}^{plateau} \times 10^{18} [cm^2]}$ & Scale factor ($s$) $\mathrm{\times 10^{17} [cm^2]}$ & Decreasing rate ($r$) & $\chi_{\nu}^2$\\
\hline
$\mathrm{H_2O}$ & 2.9 $\pm$ 0.4 & 9.7 $\pm$ 2.7 & 3.0 $\pm$ 0.2 & 4.26\\
$\mathrm{CO_2}$   & 0.3 $\pm$ 0.1 & 1.3 $\pm$ 0.4 & 1.5 $\pm$ 0.1 & 0.15\\
$\mathrm{CH_3OH^a}$ & 0.2 & 4.6 & 4.2 & 0.08\\
$\mathrm{HCOOH^a}$ & -0.03 & 0.1 & 0.64 & 0.01\\
\hline
\multicolumn{5}{l}{$^a$ Due to the poor fitting with the exponential function, the error bars of the derived parameters are now shown.}
\end{tabular}
\end{table*}

\section{MODELS OF YSOs SURROUNDED BY DUST AND ICE}

\citet{Whitney2003b} describes the physical evolution of YSOs in an evolutionary sequence, by using the 2D radiative transfer code HO-CHUNK\footnote{http://gemelli.colorado.edu/$\sim$bwhitney/codes/codes.html}. In this paper, however, the models used by Whitney et al. are employed as a template, and laboratory data of energetically processed ice are included in the simulations to address how the ice column density changes during the protostellar evolution. The physical parameters adopted in these simulations and the dust-ice model are, described in the next sections. 

\subsection{Template models}
All models share the same parameters of the central star, namely, R(R$_{\odot}$) = 2.09, T(K) = 4000 K, M(M$_{\odot}$) = 0.5 and L(L$_{\odot}$) = 1.0. A flared disk in hydrostatic equilibrium is set by the density profile below:
\begin{equation}
\rho_{disk} = \rho_0 \left(1 - \sqrt{\frac{R_{\star}}{\bar{\omega}}} \right) \left(\frac{R_{\star}}{\bar{\omega}} \right)^{\alpha} \mathrm{exp} \left\{ -\frac{1}{2} \left[\frac{z}{h(\bar{\omega})} \right]^2 \right\},
\label{rhod}
\end{equation}
where $\bar{\omega}$ is the radial coordinate in the disk midplane, and $h = h_0\left(\bar{\omega}/R_{\star} \right)^{\beta}$ is the scale height. It assumed in all models that $\alpha = 2.25$, $\beta = 1.25$ and a inner disk scale height $h_0 = 0.01$. The infalling envelope is characterized by the Ulrich's density structure \citep{Ulrich1976}, given by:
\begin{equation}
\begin{split}
\rho_{env} = \frac{\dot{M}_{env}}{4\pi} \left(\frac{GM_{\star}}{R_c^3} \right)^{-1/2} \left(\frac{r}{R_c} \right)^{-3/2} \left(1-\frac{\mu}{\mu_0} \right)^{-1/2} \\
\left(\frac{\mu}{\mu_0} + \frac{2\mu_0^2 R_c}{r}\right)^{-1},
\label{rhoenv}
\end{split}
\end{equation}
where $\dot{M}_{env}$ is the envelope mass infall rate, $R_c$ the centrifugal radius and $\mu = \cos(\theta)$. $\theta$ is the angle from the axis of symmetry and $\mu_0$ is the cosine polar angle of a streamline of infalling particles for $\mathrm{r \rightarrow \infty}$ given by
\begin{equation}
\mu_0^3 + \mu_0\left(\frac{r}{R_c} - 1\right) -\mu \left(\frac{r}{R_c} \right) = 0.
\label{particle}
\end{equation}

The evolutionary sequence is simulated by varying the parameters in Equations \ref{rhod} - \ref{particle} for the disk and envelope, as shown in Table \ref{tab:var}. The values used for each evolutionary stage were constrained from previous observational and theoretical works as pointed out by \citet{Whitney2003b}. Briefly, these models start with the central protostar surrounded by a massive envelope with a high infall rate, that decreases across the evolution. The disk mass remains constant from Class 0/I to Class II, and decreases in 6 orders of magnitude in Class III. Both inner and outer disk and envelope radius are the same as \citet{Whitney2003b}. As expected from on images of molecular outflows, the cavity angle increases with the age, whereas the cavity density decreases \citep{Padgett1999, Mottram2017, Valon2020}.

\begin{table*}
\caption{Physical parameters used in the 2D continuum radiative transfer simulation.}
\centering
\label{tab:var}
\begin{tabular}{c c c c c c }
\hline\hline
Parameter & Class 0/I & Class I & Class I/II & Class II & Class III\\
\hline
\multicolumn{1}{l}{Envelope infall rate ($\dot{M}_{env}$) [$M_{\odot} \; yr^{-1}$]} & 1 $\times$ 10$^{-5}$ & 5 $\times$ 10$^{-6}$ & 1 $\times$ 10$^{-6}$ & 0 & 0\\
\multicolumn{1}{l}{Envelope mass ($M_{env}$) [$M_{\odot}$]} & 0.37 & 0.19 & 0.037 & 1 $\times$ 10$^{-4}$ & 2 $\times$ 10$^{-5}$\\
\multicolumn{1}{l}{Envelope inner radius ($R_{e,min}$) [$R_{\star}$]} & 7.5 & 7.0 & 7.0 & 7.0 & 50.0\\
\multicolumn{1}{l}{Envelope outer radius ($R_{e,max}$) [AU]} & 5000 & 5000 & 5000 & 500 & 500\\
\multicolumn{1}{l}{Disk mass ($M_{disk}$) [$M_{\odot}$]} & 0.01 & 0.01 & 0.01 & 0.01 & 2 $\times$ 10$^{-8}$\\
\multicolumn{1}{l}{Disk inner radius ($R_{d,min}$) [$R_{\star}$]} & 7.5 & 7.0 & 7.0 & 7.0 & 50.0\\
\multicolumn{1}{l}{Disk outer radius ($R_{d,max}$) [AU]} & 50 & 200 & 300 & 300 & 300\\
\multicolumn{1}{l}{Disk accretion rate ($\dot{M}_{disk}$) [$M_{\odot} \; yr^{-1}$]} & 2.8 $\times$ 10$^{-8}$ & 6.8 $\times$ 10$^{-9}$ & 4.6 $\times$ 10$^{-9}$ & 4.6 $\times$ 10$^{-9}$ & 0\\
\multicolumn{1}{l}{Disk accretion luminosity ($L_{acc} (R_{d,min})$) [$L_{\star}$]} & 0.0069 & 0.0018 & 0.0012 & 0.0012 & 0\\
\multicolumn{1}{l}{Cavity density [$n_{H_2}$ cm$^{-3}$]} & 6.7 $\times$ 10$^{4}$ & 5.0 $\times$ 10$^{4}$ & 1.0 $\times$ 10$^{4}$ & 5.0 $\times$ 10$^{3}$ & 1.0 $\times$ 10$^{3}$\\
\multicolumn{1}{l}{Cavity opening angle [deg]} & 10 & 20 & 30 & 90 & 90\\
\hline
\end{tabular}
\end{table*}

\subsection{Dust and ice properties}
The dust properties in the models changes between the envelope, outflow cavity, upper disk layers and midplane regions. Since that physical processes such as settling and coagulation \citep{Henning2013} might take place at the midplane, millimeter-sized grains ($\sim$ 1~mm) where placed in dense regions given by $\mathrm{n_{H_2} > 10^{8} \; cm^{-3}}$. The upper disk layers ($\mathrm{n_{H_2} < 10^{8} \; cm^{-3}}$) were populated by grains with size between 1-10 $\mu$m. For the envelope, a size distribution between 0.5-1.0 $\mu$m was adopted, whereas a fixed size of 0.1 $\mu$m was used for the cavity region. It should be mentioned that all these dust models are already available in the HO-CHUNCK code.

\begin{figure*}
\includegraphics[width=18cm]{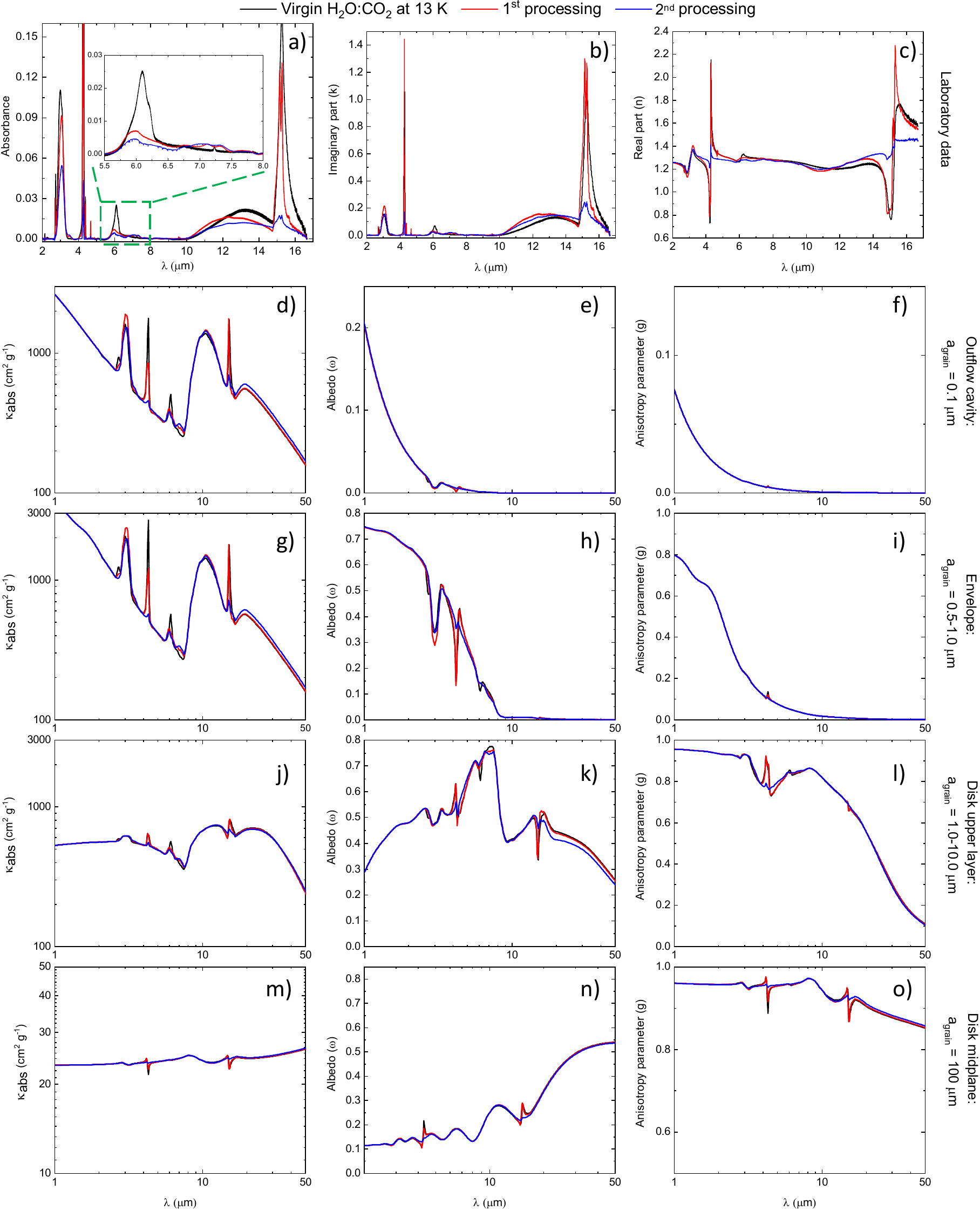}
\caption{Laboratory data and opacity models. Panels a$-$c show the absorbance and the optical constants {\it n} and {\it k} at different fluences. The inner figures in panel a show different regions of the absorbance spectrum. Panels d-o show the dust parameters used in the models where the temperature is below 150~K, and for different regions in the YSOs (See details in the text).}
\label{fig:Opac}
\end{figure*}

The ice mantle is the new part in these models, compared to the original paper. Although the authors have also used H$_2$O ice in their models, this paper employs an ice composition given by the mixture of H$_2$O:CO$_2$ (1:1) processed by $^{58}$Ni$^{13+}$ ions from its pristine composition: (i) Fluence 0 (Virgin ices), (ii) Fluence 1$\times$10$^{12}$ ions cm$^{-2}$ (first processing) and (iii) Fluence 1$\times$10$^{13}$ ions cm$^{-2}$ (second processing) obtained from \citet{Pilling2010} (see Figure \ref{fig:Opac}a). As calculated by \citet{Drury1999} and \citet{Shen2004}, the $^{58}$Ni flux is around 6.4 $\times$ 10$^{-6}$ cm$^{-2}$ s$^{-1}$, which corresponds to a relative time to the virgin ice of of 5 $\times$ 10$^5$ year and 5 $\times$ 10$^6$ year for the Fluences 1 and 2, respectively. These two irradiated ice spectra were selected from the whole range of experiments because they cover the timescales of Class I and Class II YSOs. The inner panels in Figure \ref{fig:Opac}a show the ice features that will be used to calculate the ice column densities, except the range between 14$-$17 $\mu$m. It is important to note from the shape of the O-H stretching mode that no water crystallization is induced by the CR processing, although the ice segregation is evident from the formation of the double peak of CO$_2$ feature at 15~$\mu$m. 

Although the CO$_2$/H$_2$O ratio is expected to be around 30\% toward low-mass YSOs \citep{Oberg2011}, a higher ratio is used in this paper. Even though this high CO$_2$ fraction is unlikely for the star-forming regions, it was used in the experiments to maximize the formation of C-bearing products. However, as a product of the ice processing in the experiments, carbon monoxide ice was only detectable in the laboratory spectrum after the first irradiation dose, thereby being destroyed at high fluences. Given this unique detection of CO ice in the selected experiments used in this paper, its chemical evolution with the evolutionary stage of YSOs was not addressed in this paper. Another sample containing CO in the initial mixture could be also explored. The effective interaction limit of other ionizing agents as UV and X-rays is much lower compared to Cosmic Rays \citep{Indriolo2013} in dense regions such as the protostellar envelope. Cosmic ray-processing, on the other hand, might remove electrons from the inner shells of atoms, or excite H$_2$ molecules, leading to an induced X-ray and/or UV radiation field.

The optical constants of the H$_2$O:CO$_2$ ices were taken from \citet{Rocha2017} (Figure \ref{fig:Opac}b-c). In order to create an ice-covered dust model for the envelope and midplane, the Maxwell-Garnett effective medium theory \citep{Bohren1983} and the Mie theory were employed to calculate the absorption opacity, albedo ($\omega$) and anisotropy parameter (g), as shown in Figure \ref{fig:Opac}d-o for the outflow cavity, envelope, disk upper layer and disk midplane.

The position of the ices in the disk and envelope was defined via an interactive procedure by calculating the dust temperature using the Monte Carlo method. As the first approach, the dust temperature was calculated for all models, but without ice components. The calculated temperature was used to replace bare grains by covered grains, using the water desorption temperature. Since the new temperature distribution is different compared to the models using bare grains, the simulations were repeated until the temperature converges for values around 5\% of difference between the prior and posterior calculations, as also done in \citet{Pontoppidan2005}.

Three models of YSOs including ices are addressed in this paper. The virgin ice model describes a physical evolution without chemical evolution since the ices are not processed during the evolutionary sequence. 1$^{st}$ processing model assumes an ice mantle slightly processed by external Cosmic Rays (CRs), whereas the 2$^{nd}$ processing Model 3 is given by ices highly processed by CRs.

\subsection{Density and Temperature profiles}
\begin{figure*}
\includegraphics[width=\textwidth,height=\textheight,keepaspectratio]{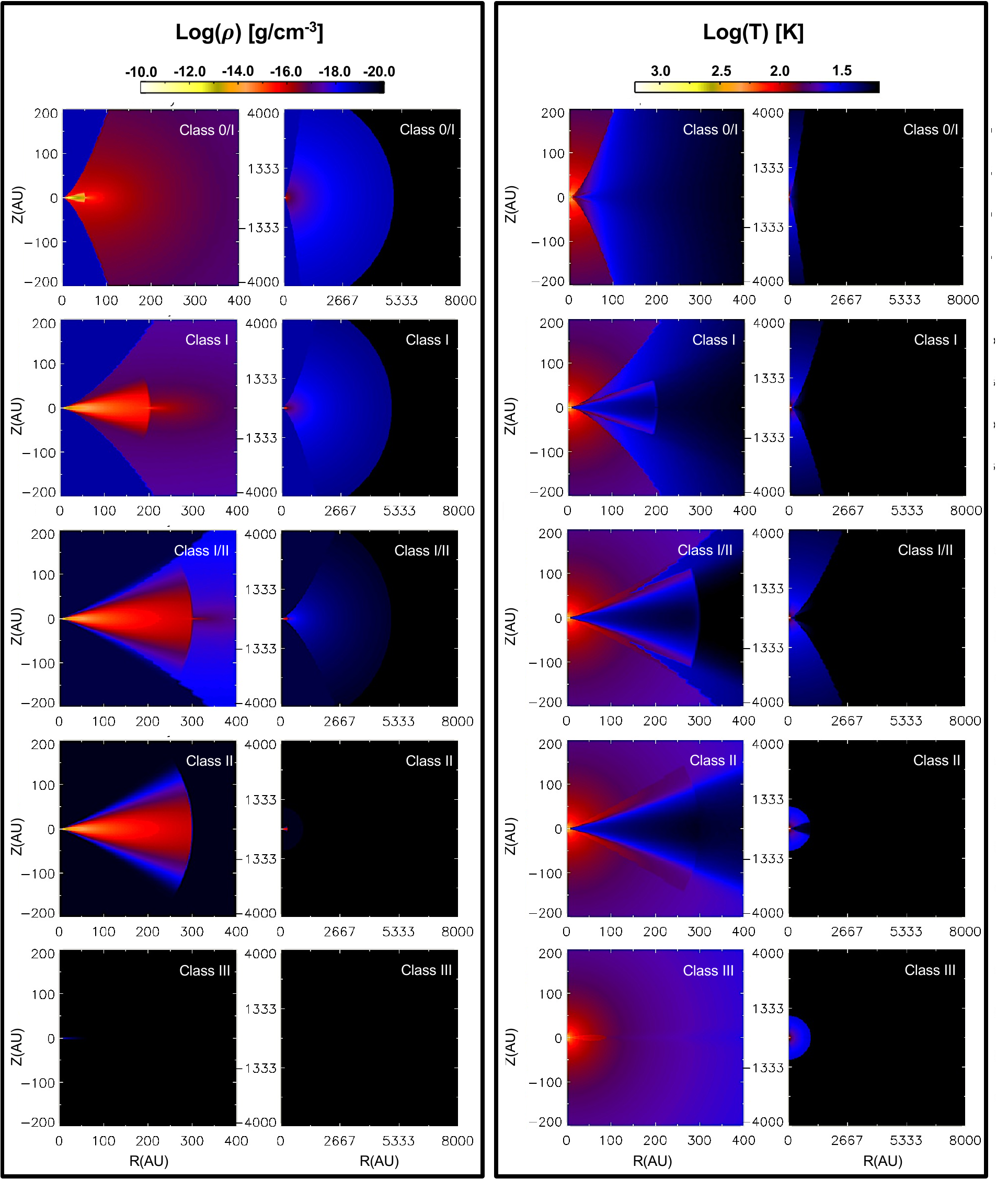}
\caption{Temperature and density profile calculated with the radiative transfer code for the Class 0/I - III. The colours are shown in logarithmic scale and the values are indicated by the bar on the top. Z and R are, respectively, the radial and vertical scale, both given in astronomical units (AU). For each box, the left models are a zoom-in of the right ones.}
\label{fig:T_D}
\end{figure*}

Figure \ref{fig:T_D} show the temperature and density distribution for the evolutionary sequence between Class 0/I and Class III. The left panels in each box is a zoom-in of the right ones. The left box displays the density distribution that ranges between $\mathrm{10^{-18.5}~-~10^{-19.5}~g~cm^{-3}}$, i.e $10^4$ $-$ $10^{5}$ cm$^{-3}$ in the Class 0/I and I and is extended to large scales ($R_{\mathrm{out}}$ = 5000~AU). In Class II, this density variation is distributed in a disk scale until 300~AU. The right box shows the dust temperature distribution in each model, ranging from T $\sim$ 1000~K nearby the central star to T $\sim$ 10$-$30~K at the envelope region. At such a lower temperature, the outer envelope still cold enough to host ice-covered dust grains, which can be processed by the Interstellar Radiation Field (ISRF). For very embedded YSOs, the UV field might not dominate the synthesis of molecules, unless it is 40 times higher than the typical ISRF of the Interstellar Medium \citep{Rocha2018}. In that case, the chemistry is dominated by the low-temperature reaction, such as the cosmic ray induced processes.

In the disk region, the temperature in the midplane remains at around 15$-$20~K in all evolutionary stages due to UV shielding caused by the dust grains. In the upper disk layers, on the other hand, the temperature increases as the envelope mass decrease because the accretion process, and the temperature is around 100~K for Class I and 300~K for Class II at a radius of 100~AU. As a consequence, the major part of the Class I disk might host ices since the H$_2$O freeze-out occurs at temperatures below 150~K \citep{Collings2004}. The ice reservoir in Class II disks is therefore reduced compared to the previous evolutionary stage.

\begin{figure}
\includegraphics[width=8.5cm]{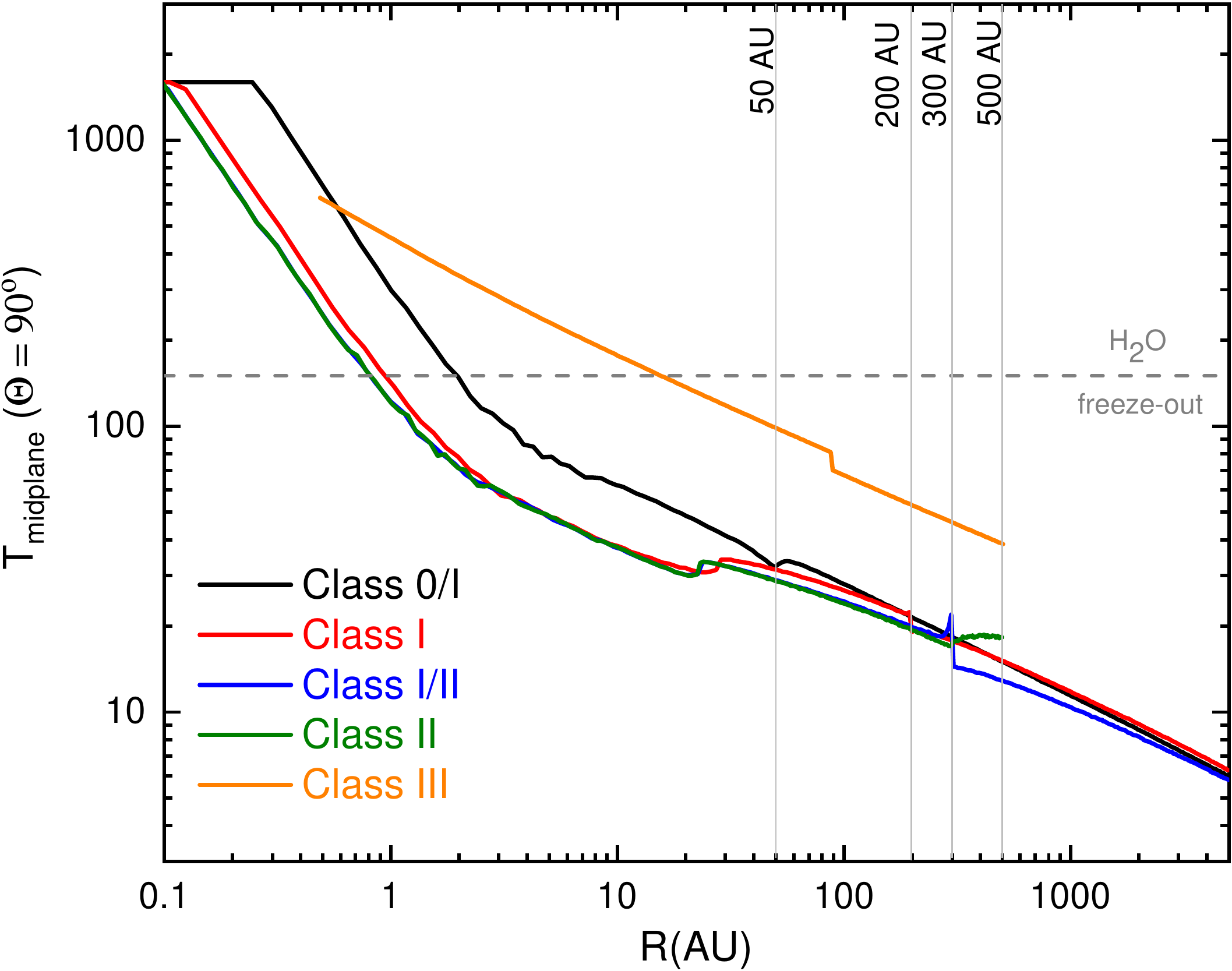}
\caption{Radial dust temperature at $\mathrm{\theta = 90^{\circ}}$ of the models in different stages, as indicated by the colour scheme. The water freeze-out line is also shown by the dashed grey line. The vertical grey lines indicate the outer radius of the disk (50~AU, 200~AU, 300~AU) and the outer envelope radius in evolved stages (500~AU).}
\label{fig:Tdust}
\end{figure}

Figure \ref{fig:Tdust} shows the dust temperature across the midplane for all the stages, and the H$_2$O freeze-out limit is indicated. The models show that water snowline moves inward until the Class II, and outwards for the Class III, which is in agreement with previous works \citep{Kennedy2008, Baillie2015}, and would benefit the formation of giant planets at around 5~AU without significant migration. Quantitatively, \citet{Kennedy2008} and \citet{Baillie2015} suggest that the water snowline changes from 3.5~AU (Class 0/I) to 1.7~AU (Class II) and from 2.3~AU to 1.5~AU (same stages), respectively. In this work, the snowline is closer to the central star, compared to the previous models since it changes from 2~AU in Class 0/I to 0.9~AU in Class II. Despite the number difference, the same trend and the reduction by a factor of 2 are still observed. It worth to note, however, that snowlines might temporarily move outwards in the disk in a scenario of inside-out collapse \citep{Zhang2015, Cieza2016}.

\subsection{Spectral Energy Distributions}
\begin{figure*}
\includegraphics[width=\textwidth,height=\textheight,keepaspectratio]{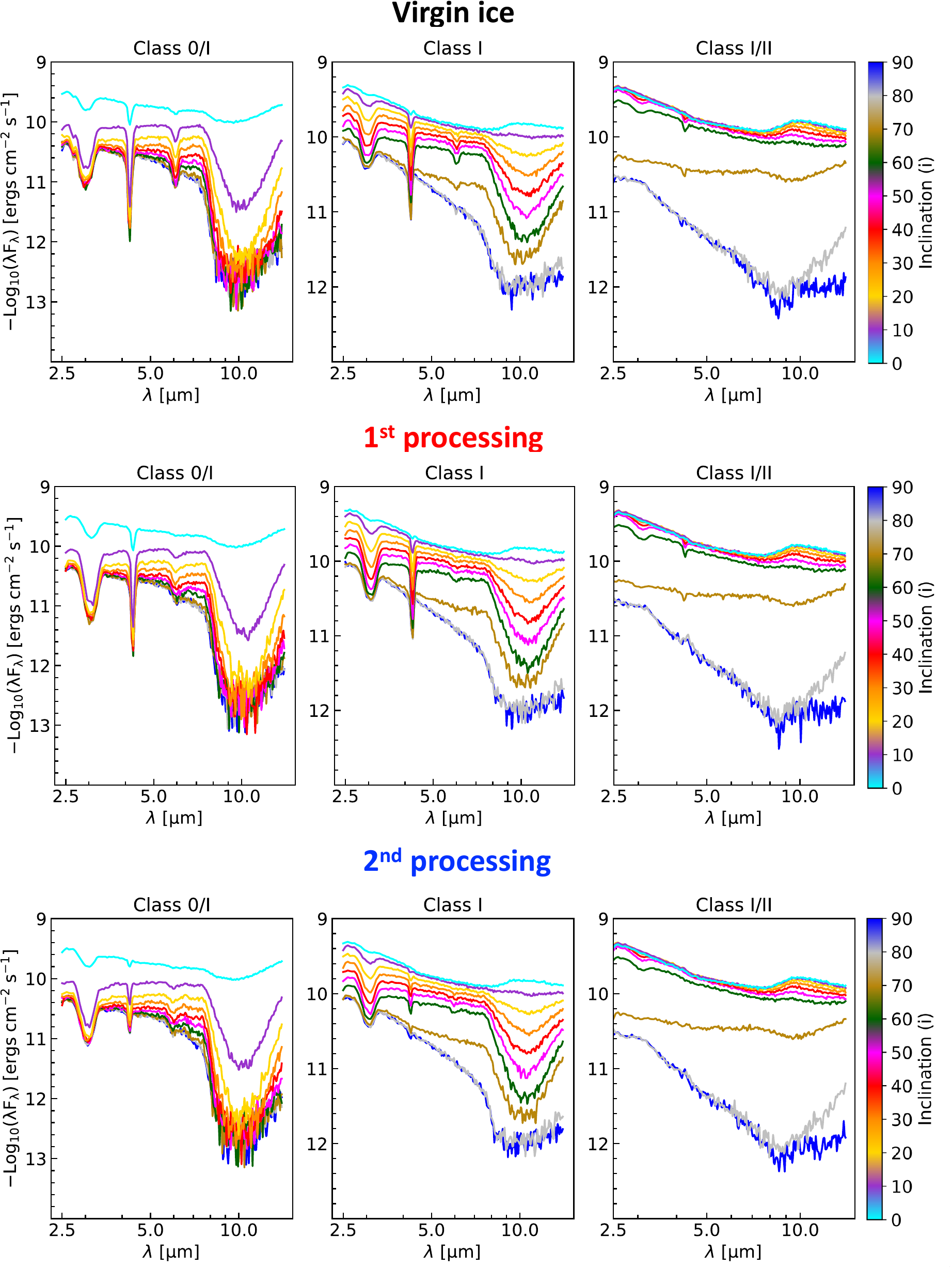}
\caption{Spectral Energy Distribution of the YSOs simulated in this paper for the Class 0/I$-$I/II at three levels of ice processing and 10 inclinations.}
\label{fig:SEDg}
\end{figure*}

Figure \ref{fig:SEDg} shows the SEDs of the simulated protostars at 10 inclination angles ($i$). Since we are looking at the ice features between 2.5$-$8.0 $\mu$m, a small range of wavelengths is shown, although the radiative transfer simulation was performed for 0.1$\mu$m $\leq$ $\lambda$ $\leq$ 1000 $\mu$m. In order to get the best signal-to-noise ratio due to the Monte Carlo noise in the simulation of dense regions, a number of 4 $\times$ 10$^7$ photons was used in the models. The SEDs show strong absorption bands which are associated to the H$_2$O ice (3~$\mu$m and 6~$\mu$m), CO$_2$ ice (4.27~$\mu$m) silicate (9.8~$\mu$m) and less noticeable in this scale shown, the contribution of complex molecules between 5.5$-$8.0~$\mu$m in the 1$^{st}$ and 2$^{nd}$ processing panels (see Figure~\ref{fig:tauCOM}). The silicate feature is seen in absorption for all inclinations in the Class 0/I stage, whereas it turns to an emission profile at a pole-on inclination in Class I due to the low optical depth to the source of emission. Such an emission profile is more pronounced for $i \leq 60^{\circ}$ at Class I/II since the envelope mass decreases by a factor of 5 from the previous to the new stage in these models. The ice profiles in the Class 0/I and Class I, are always observed in absorption at all inclinations. In the Class I/II, on the other hand, the ice absorption is only seen between $i = 20^{\circ}$ and $i = 60^{\circ}$. In extremely edge-on ($i = 80^{\circ}$ and $i = 90^{\circ}$) and face-on ($i = 20^{\circ}$ and $i = 10^{\circ}$) inclinations, the ice features are weak. It worth to note that the intensity of the ice absorption does not vary monotonically with the viewing angle, namely, from edge-on to face-on inclination as will be discussed in Section \ref{Res_Diss}.

\begin{figure*}
\includegraphics[width=\textwidth,height=\textheight,keepaspectratio]{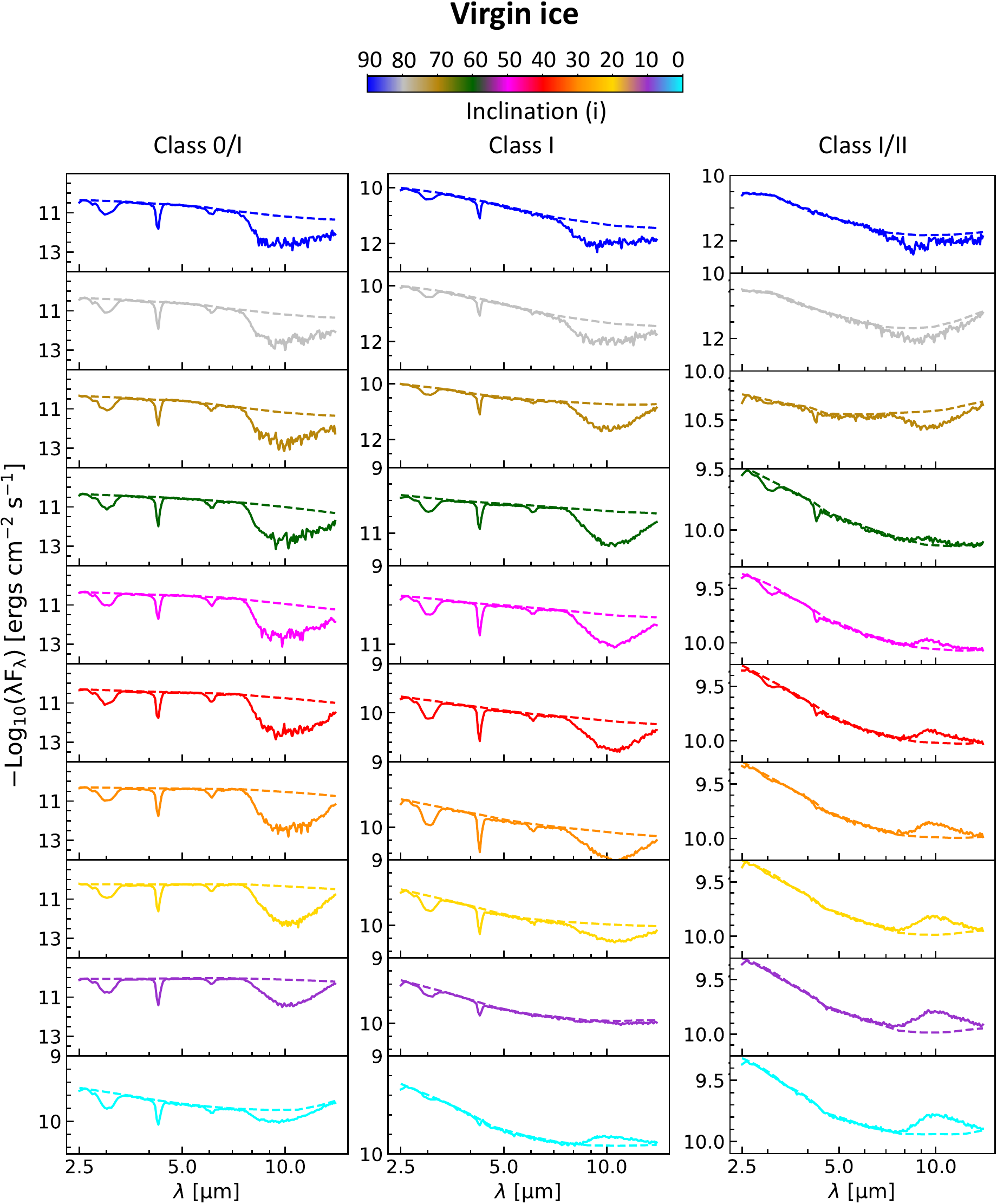}
\caption{Spectral Energy Distribution for the model of virgin ices, where the continuum line is also indicated by the dashed lines. The colours indicate the inclinations.}
\label{fig:SEDV}
\end{figure*}

In addition to the SEDs, the continuum emission was also calculated for all 90 spectra described here. As discussed in \citet{Boogert2008}, determining the best baseline is not trivial in real objects, and very accurate modelling is required to determine how the different regions of the disk or disk+envelope contribute to the continuum emission. In this paper, however, all SED components are known and were used to calculate the correct baseline for each model. For instance, Figure \ref{fig:SEDV} show the continuum given by the dashed line for SEDs in the model of virgin ices in Figure \ref{fig:SEDg}. The same procedure was employed for the models of 1$^{st}$ and 2$^{nd}$ processing.

\section{Results and discussion}
\label{Res_Diss}
\subsection{Ice optical depths}
\begin{figure*}
\includegraphics[width=\textwidth,height=\textheight,keepaspectratio]{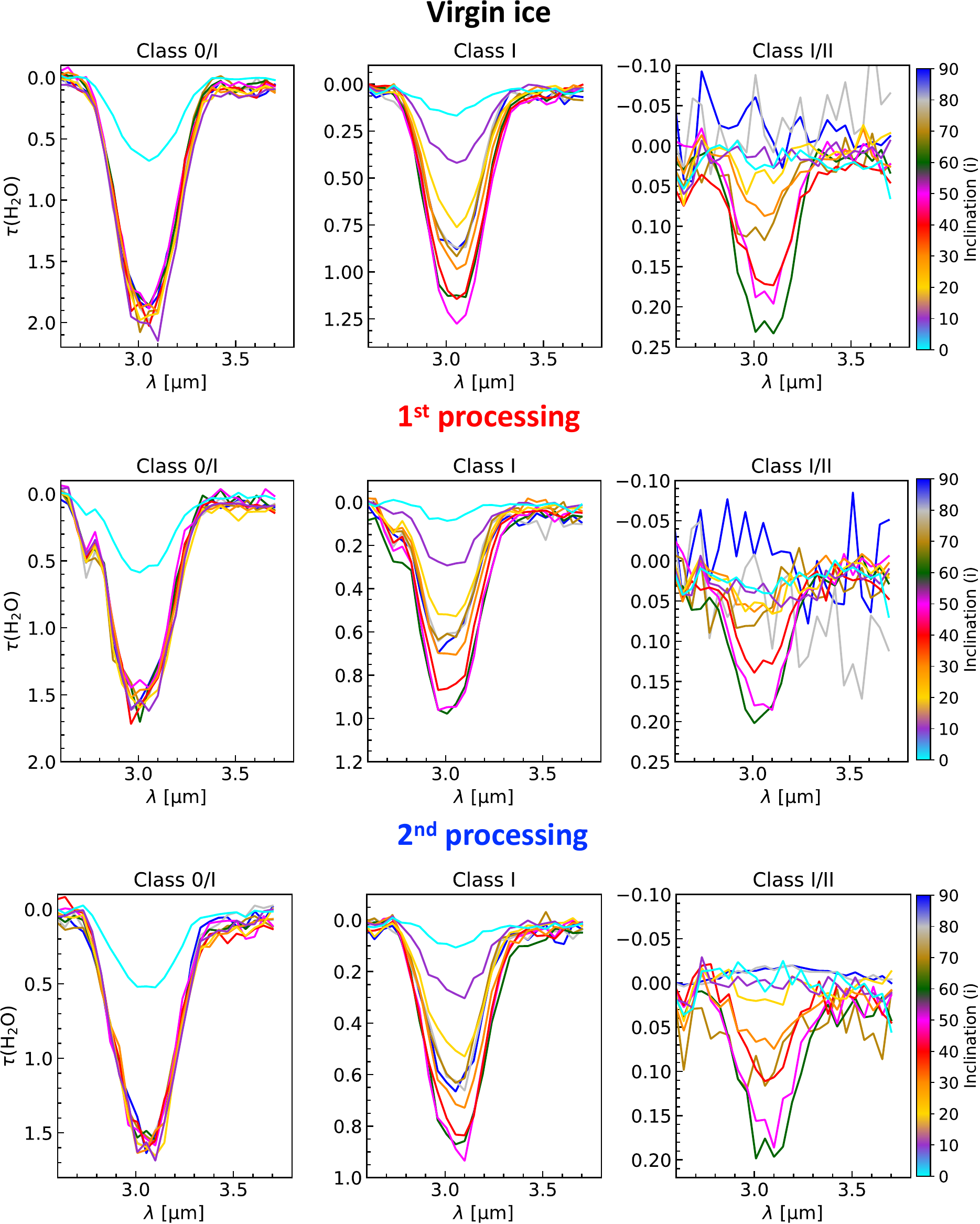}
\caption{Optical depth of H$_2$O ice at around 3 $\mu$m shown for the Class 0/I$-$I/II at three levels of ice processing and 10 inclinations as indicated by the colour bar.}
\label{fig:tauH2O}
\end{figure*}

\begin{figure*}
\includegraphics[width=\textwidth,height=\textheight,keepaspectratio]{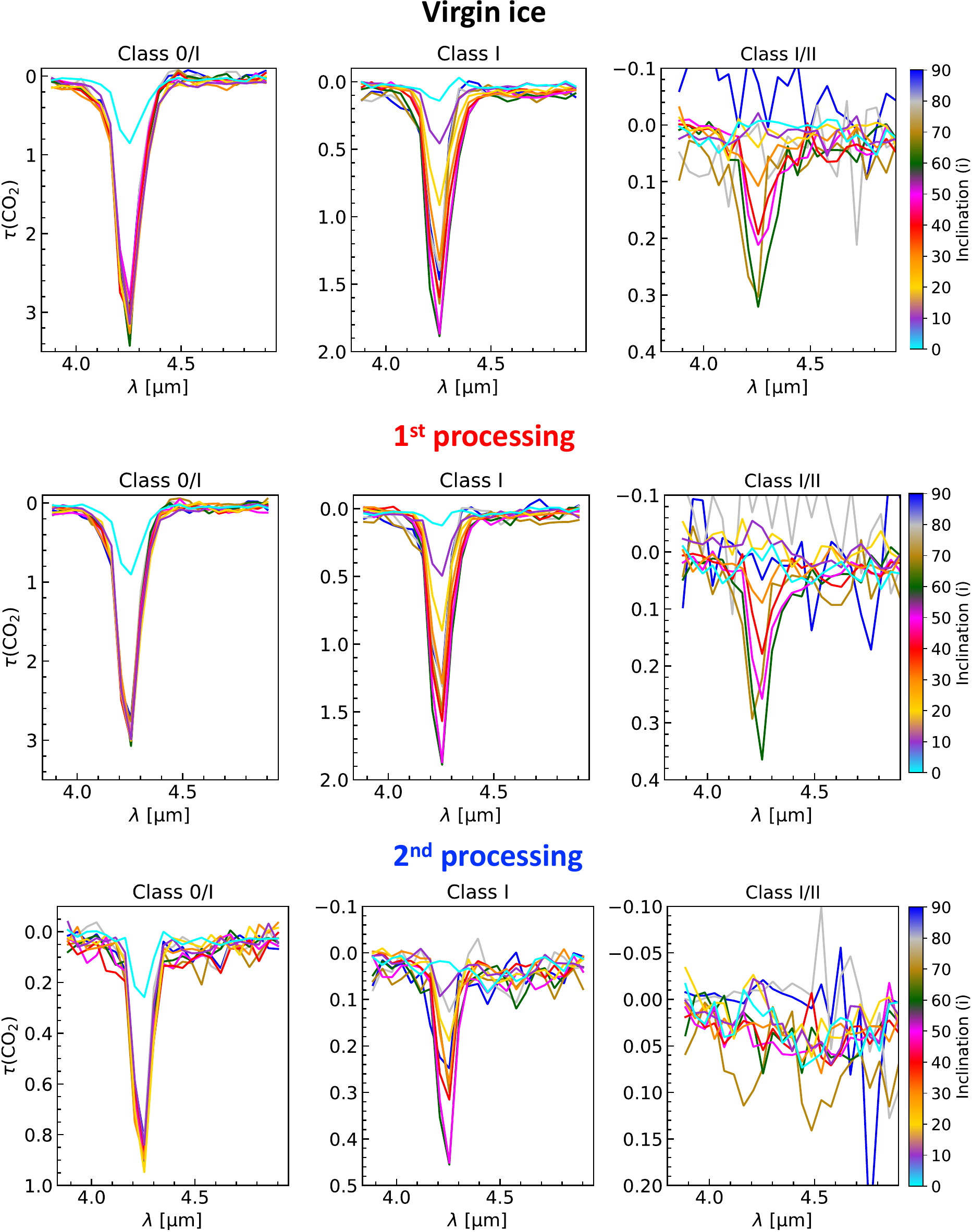}
\caption{Optical depth of CO$_2$ ice at around 4.27~$\mu$m shown for the Class 0/I$-$I/II at three levels of ice processing and 10 inclinations as indicated by the colour bar.}
\label{fig:tauCO2}
\end{figure*}

\begin{figure*}
\includegraphics[width=\textwidth,height=\textheight,keepaspectratio]{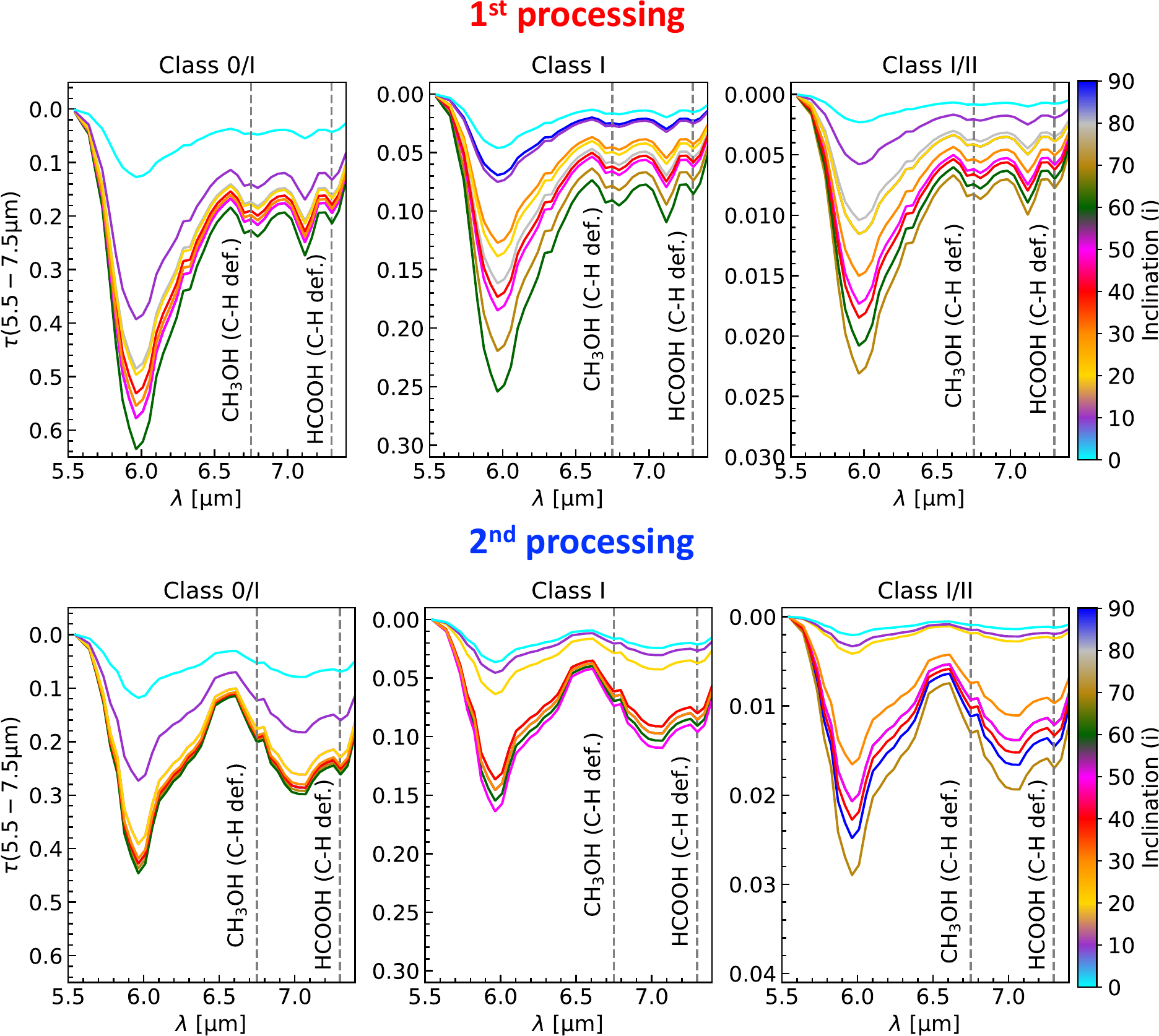}
\caption{Optical depth between 5.5$-$7.5~$\mu$m shown for the Class 0/I$-$I/II at two levels of ice processing and 10 inclinations as indicated by the colour bar. The contributions of CH$_3$OH and HCOOH are indicated by the vertical and grey dashed lines at 6.75 $\mu$m and 7.24 $\mu$m, respectively.}
\label{fig:tauCOM}
\end{figure*}

The Optical Depth ($\tau$) for the models of virgin and processed ices, related to H$_2$O ice ($\sim$ 3 $\mu$m), CO$_2$ ice ($\sim$ 4.27 $\mu$m) and the complex molecules between 5.5$-$7.5~$\mu$m were calculated at 10 inclinations between face-on and edge-on angles using the equation described in Section~2. Figures \ref{fig:tauH2O}, \ref{fig:tauCO2} and \ref{fig:tauCOM} show the ice features for each case. The optical depth variation with inclination is due to the relation between optical depth and density along the line of sight. One can note from Equation \ref{rhoenv} that $\rho_{env}$ decreases with the radius $r$, but also with the $\cos(\theta)$, namely, from the midplane toward the cavity region. Since the optical depth is given by $\mathrm{d\tau} = -\kappa \rho \cos{\theta} ds$, where $\kappa$ is the opacity, $\rho$ the density and $ds$ is the optical path, $d\tau$ decreases if $\rho_{env}$ also does. Furthermore, the small variation observed in Class 0/I compared to Class I and Class I/II is due to the negligible envelope density variation with the polar angle for small cavity apertures. In Class I and Class I/II, however, the ice optical depth does not vary monotonically between $i = 90^{\circ}$ and $i = 60^{\circ}$. It is evident that the inclination of $i = 70^{\circ}$ shows the deepest ice bands due to a geometric effect as previously shown by \citet{Pontoppidan2005}. Due to the large optical depth through the midplane, the IR source at edge-on inclination is dominated by a small fraction of scattered infrared photons toward high angles. However, at an inclination above the disk opening angle, the stellar emission itself dominates and the entire envelope is probed, leading to deepest ice bands in the spectrum. Below $i = 70^{\circ}$ in the models shown in this paper, the ice column density decreases monotonically until pole-on inclinations.

The H$-$O stretching vibrational mode seen in Figure \ref{fig:tauH2O} between 2.7 and 3.6 $\mu$m is usually reported in YSOs as containing a red wing at longer wavelengths caused by scattering on large ice-coated grains \citep{Boogert2000} and the absorption by ammonia hydrates \citep{Hagen1983}. The absence of this effect in the synthetic spectra does not argue in favour of any of these two cases since neither large grains or N-containing species were included in the ice-dust model for the envelope. It is also true that this water vibrational mode shows evidence of crystallization if the ice is heated above 100~K. However, as noticed from Figure \ref{fig:Opac}a, the CR-processing is not enough to induce discernible structural changes in the ice matrix as seen from the IR spectrum.

In Figure \ref{fig:tauCO2}, the C$-$O stretching mode of CO$_2$ ice at around 4.27~$\mu$m is shown. As discussed in \citet{BoogertCDE,Ehrenfreund2001} the position and width of this peak changes with the grain geometry and its fraction in the ice matrix. As pointed out by Ehrenfreund et al., this vibrational mode presents a narrow profile if less abundant in a polar matrix compared to the pure ice. For instance, the fraction of 14\% of CO$_2$ ice in a H$_2$O matrix fits better the CO$_2$ band at 4.27~$\mu$m for the Class I YSO Elias~29 compared to the pure CO$_2$ ice adopted by \citet{Rocha2015}. The CO$_2$ analysis at $\sim$ 15 $\mu$m from {Spitzer} observations \citep{Pontoppidan2008} has strongly suggested that 2/3 of the CO$_2$ absorption features is due to carbon dioxide ice diluted in H$_2$O ice, whereas 1/3 is likely due to CO:CO$_2$ mixture. \citet{Ehrenfreund1997} shows a narrowing of the CO$_2$ absorption feature if mixed in CO ice at fractions below 26\%. As a consequence, if the band strength ($\mathcal{A}$) is kept constant, this effect would lead to lower column density values since it depends on the integrated optical depth as seen in Equation~2. Nevertheless, this is hard to verify since the $\mathcal{A}$ for CO$_2$ diluted in CO ice is unknown.

Figure \ref{fig:tauCOM} shows the optical depth between 5.5$-$7.5 $\mu$m that usually is associated with the presence of Complex Organic Molecules such as HCOOH, CH$_3$OH, CH$_3$CHO, CH$_3$CH$_2$OH \citep{Schutte1996, Gibb2000, Keane2001, Boogert2008, Boogert2015}. In this work, we highlight the contribution of methanol and formic acid at 6.75~$\mu$m and 7.24~$\mu$m, respectively, as a result of the ice processing by cosmic rays. Addressing the ice components in this spectral range is rather difficult since the IR profile of complex molecules changes with the chemical environment inside the ice matrix. \citet{Boogert2008} suggests a decomposition method by removing the pure water contribution, and fitting the residual with 5 independent absorption components obtained from the combination of different YSO spectra. Each component might have multiple carriers due to blended vibrational modes, as for example CH$_3$OH and NH$_4^+$ at 6.85~$\mu$m. In this paper, however, a simple Gaussian decomposition of the spectrum by using 5 components is employed in order to isolate the contribution of CH$_3$OH and HCOOH, whose the position in spectra are shown in \citet{Pilling2010}. The Appendix A shows the Gaussian decomposition method, where the components due to CH$_3$OH and HCOOH ice are shown.

\subsection{Synthetic column density and the spectral index}
In order to avoid blending effects, the H$_2$O and CO$_2$ column densities were calculated from the bands at 3~$\mu$m and 4.27~$\mu$m shown in Figures \ref{fig:tauH2O}-\ref{fig:tauCOM}, using Equation \ref{CD}. In the case of  CH$_3$OH (6.85~$\mu$m) and HCOOH (7.25~$\mu$m), the spectral decomposition method described in Appendix~A was used. The band strength adopted for each vibrational mode were: $\mathcal{A_{\mathrm{H_2O}}}$~=~2.6 $\times$~10$^{-16}$ cm molecule$^{-1}$, \citep{Hagen1981}, $\mathcal{A_{\mathrm{CO_2}}}$~=~7.6 $\times$~10$^{-17}$ cm molecule$^{-1}$ \citep{Gerakines1995}, $\mathcal{A_{\mathrm{HCOOH}}}$~=~1.5 $\times$~10$^{-17}$ cm molecule$^{-1}$ \citep{Park2006}, and $\mathcal{A_{\mathrm{CH_3OH}}}$~=~1.2 $\times$~10$^{-17}$ cm molecule$^{-1}$ \citep{Hudgins1993}. This paper assumes a constant $\mathcal{A}$ for the 3 models, although the band strength is a sensitive physicochemical parameter to the ice composition. As shown by \citet{Oberg2007}, the band strength of the H$_2$O bulk stretch at 3~$\mu$m drops linearly in H$_2$O:CO$_2$ ice mixture, compared to the pure H$_2$O. However, in the scenario of energetic processing, where several species are formed, determining an accurate band strength still an open problem in astrochemistry. 

\begin{figure*}
\includegraphics[width=\textwidth,height=\textheight,keepaspectratio]{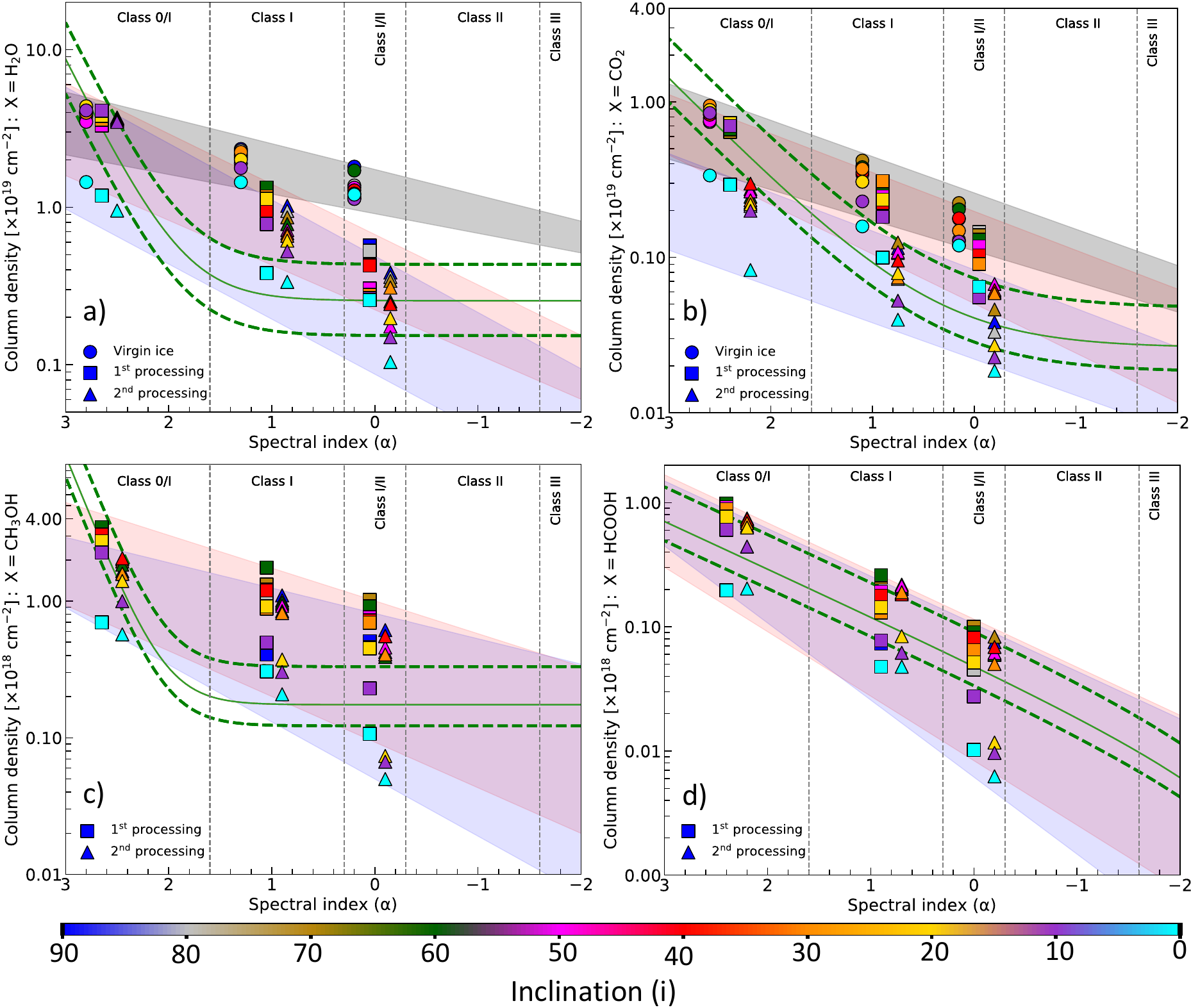}
\caption{Plot containing the spectral index against the column density of the ices H$_2$O, CO$_2$, CH$_3$OH and HCOOH calculated from the models of virgin and processed ices. The three levels of ice processing are indicated by the filled circles, squares and triangles, respectively. The inclination effect is shown by the color scheme given by the colorbar. Each evolutionary stage is indicated by the top labels. The solid and dashed green lines correspond to the exponential fit and confidence intervals in Figure~\ref{fig:abs_alpha}. The grey, red and blue shaded regions cover column density dispersion for all inclination according to the spectral index.}
\label{fig:SyntN}
\end{figure*}

Figure~\ref{fig:SyntN} shows the synthetic column density ($N$) of the ices $X$: H$_2$O, CO$_2$, CH$_3$OH and HCOOH as a function of the spectral index calculate for the model of virgin ices, 1$^{st}$ processing and 2$^{nd}$ processing shown by the filled circles, squares and triangles, respectively. The colours indicate the inclination angle as given by the colorbar at the bottom of the figure. The virgin ice model represents a scenario where the ice column density varies only due to the envelope dissipation and angle inclination. The two processing models, on the other hand, simulate a case where the ice column density decreases due to the physical effects in the virgin ice model, as well as due to the ice energetic processing. Due to the ambiguity in the envelope mass with the spectral index shown by \citet{Crapsi2008}, the spectral index ($\alpha$) to characterize the evolutionary stage of YSOs in this paper is kept constant for each evolutionary stage and processing level. For Class 0/I, the adopted $\alpha$ values are 2.8, 2.6 and 2.4 for the virgin ice, 1$^{st}$ and 2$^{nd}$ processing, respectively. The small offset is only to allow better readability. At Class I, the $\alpha$ values are 1.4, 1.1 and 0.9, whereas for Class I/II, it is assumed 0.2, 0, -0.2. In this way, Figure~\ref{fig:SyntN} shows the comparison between model and observation from the perspective of the inclination angle and ice processing level. As pointed-out in Section~4.4, the ice column density does not vary monotonically with the 10 inclination angles shown in Figure~\ref{fig:SyntN}. The maximum N$_X$ occurs for $60^{\circ} \leq i \leq 70^{\circ}$, whereas the minimum N$_X$ corresponds to $0^{\circ}$. The observational fit and the confidence intervals taken from Figure~\ref{fig:abs_alpha} are shown by the solid and dashed green lines, respectively. The grey, red and blue shaded areas cover the ice column density dispersion for the three models used in this paper according to the spectral index.

Figure~\ref{fig:SyntN}a shows that, except for the pole-on inclination angle, the H$_2$O column densities predicted by the three models lie inside the confidence interval of the observational fit at the stage of Class 0/I objects. However, the inclination effect and ice processing cannot explain the vertical spread of two orders of the magnitude observed at this stage. Even though the pole-on column densities are 50$-$70\% lower than the highest inclinations, they are still unable to explain the entire vertical variation. A likely cause of this wide vertical spread in the same evolutionary stage is the initial envelope mass variation associated with protostars at early stages, since that, at the same physical conditions, lower envelope mass leads to a lower ice column density. At the Class I stage, all the 3 models overestimate the water ice column density, except for the pole-on inclination, which suggests that H$_2$O is destroyed by mechanisms other than cosmic rays radiolysis (H$_2$O~+~CR $\displaystyle \rightarrow$ OH~+~H), such as photolysis induced by UV and X-rays. At Class I/II stage, on the other hand, only the virgin ice model overestimate the H$_2$O ice column density, whereas the 1$^{st}$ and 2$^{nd}$ processing model are in agreement with the observation. Nevertheless, it is inconclusive that such a result is related to the physical and chemical evolution of the protostar since only one stage of evolution fit the data.

The trends shown by the shaded areas in Figure~\ref{fig:SyntN}a indicate a slow $N_{\mathrm{ice}}$ decreasing at early stages and the absence of the {\it plateau} for late stages when compared to the observational data. While the fast decreasing observed at early stages for water ice is likely related to other destruction mechanisms not included in this paper, the absence of {\it plateau} at late stages can suggest that, at least 30\% of the observed ices lying outside the protostar itself, and therefore, not affected by the accretion process, might contribute to keeping the observed ice column density fraction roughly constant as shown by \citet{Pontoppidan2005}, \citet{Boogert2000} and \citep{Aikawa2012}.

In the case of the $N_{\mathrm{CO_2}}$ variation shown in Figure~\ref{fig:SyntN}b, the virgin ice and the 1$^{st}$ processing models provides column densities inside the observational confidence interval in Class 0/I. The CO$_2$ destruction in the 2$^{nd}$ processing model, on the other hand underestimate $N_{\mathrm{CO_2}}$. Nevertheless, in the more evolved stages, namely, Class I and Class I/II, the 2$^{nd}$ processing model predicts CO$_2$ ice column densities inside the confidence interval, whereas the virgin ice and the 1$^{st}$ processing models, generally overestimate the column densities. As in the case of H$_2$O ice, the effect of the physical and chemical evolution of protostars cannot be confirmed. The decreasing trends shown by the shaded areas also indicate the absence of the {\it plateau} at late stages. For the Class I YSOs, Elias~29 and CRBR2422.8$-$3423, the estimated column density in foreground clouds are $N_{\mathrm{CO_2}} \sim 2 \times 10^{17}$ cm$^{-2}$ \citep{Boogert2000} and $N_{\mathrm{CO_2}} \leq 6 \times 10^{17}$ cm$^{-2}$ \citep{Pontoppidan2005}, respectively.  

Figure~\ref{fig:SyntN}c shows the variation of the methanol ice column density against the spectral index as calculated from the band at 6.85~$\mu$m spectral feature. One can note that the methanol column densities calculated from 1$^{st}$ and 2$^{nd}$ processing models mostly lie inside the observational range for Class 0/I, although the vertical spread is not explained, whereas it is overestimated in the later stages. The reason of the overestimation is likely due to (i) the high efficiency of CH$_3$OH production because of the large CO$_2$ fraction in the experiments and (ii) due to the decomposition method itself. In the latest case, \citet{Boogert2008} the component attributed to CH$_3$OH ice has a full width at half maximum (FWHM) 50\% lower than the assumed in this paper, which would lead to a lower column density calculation. Taking this difference into account, if the FWHM Gaussian component associated with methanol in this paper, is reduced in 50\%, it would represent an ice column density 20\% lower. Such a reduction, however, would not provide a better agreement between observation and model. In summary, the model fails to reproduce the methanol ice column density variation and the fast decreasing compared to the other molecules as pointed-out at the beginning of the paper. Further investigation must be carried out for this case, and other laboratory ice samples can be used instead, as well as how the methanol formation is affected during the protostellar collapse.

HCOOH ice column density variation is shown in Figure~\ref{fig:SyntN}d as calculated from the absorption feature at 7.24~$\mu$m. Both 1$^{st}$ and 2$^{nd}$ processing models provide $N_{\mathrm{HCOOH}}$ inside the observational confidence interval in most of the cases. The good agreement between observation and the two models suggest that the lower decreasing rate of formic acid is mainly due to the envelope dissipation by accretion than due to the energetic processing of the ice i.e. chemical evolution. The absence of the {\it plateau} at late stages, on the other hand, indicate that, if it is caused by ices in foreground clouds, then its abundance quiescent molecular clouds is very low.

In summary, the evolutionary models along with the ice samples used in this paper, in general, do not explain the observed trends of ice column density with the evolutionary stage, with the exception of HCOOH. Further investigations addressing different physical parameters and ice mixtures must be used in order to explain the observational data.

\section{Conclusions}
This paper shows the correlation between the ice column densities of H$_2$O, CO$_2$, CH$_3$OH and HCOOH and spectral index ($\alpha$) of 27 early-stage YSOs. A computational simulation combining 2D radiative transfer model and laboratory data of virgin and processed ices was carried-out to study the ice column density decreasing from early to late stages of the protostellar evolution. The conclusions are summarized below:

\begin{enumerate}
\item The observational data suggest that ice column densities are correlated with the spectral index from Class 0/I to Class I/II. Additionally, it is observed that H$_2$O and CH$_3$OH ices reach an ice column density {\it plateau} at Class I whereas for CO$_2$ ice, the {\it plateau} is reached at Class I/II. In the case of HCOOH, however, it is not observed.  If an exponential function is assumed to fit the data, it is found that  CH$_3$OH ice decreases faster then the other ices, whereas HCOOH decreases slowly.

\item In agreement with previous works, the models simulated in this paper, shows that the H$_2$O snow line moves inward from 2~AU at Class 0/I to 0.7~AU at Class II. During the evolution to Class III, the snowline moves outward to 20~AU. Nevertheless, the region where the ices are located is optically thick in the mid-IR, and therefore the absorption seen in the spectra are due to the ices located in the envelope. As a consequence of the density distribution in the Class 0/I to Class I/II, the ice column density does not vary monotonically with the inclination angle ($i$). The highest column density is, generally, seen for $i \sim 20^{\circ}$ at Class 0/I and $i \sim 60^{\circ}$ at Class I.

\item The computational models show that the combination of physical evolution (envelope dissipation), chemical evolution (ice processing) and inclination angle effect is able to reproduce the $N_{\mathrm{HCOOH}}$ decreasing with the spectral index. However, in the case of H$_2$O, CO$_2$ and CH$_3$OH, the models fail to reproduce the observations in all evolutionary stages. Additionally, other destruction pathways not included in the current method must be addressed to understand why water and methanol ice column density decrease faster than CO$_2$ and HCOOH.

\item The absence of the ice column density {\it plateau} in the models suggest that there is fraction of ice absorption located in foreground clouds in order to explain the observations. More accurate models including other physical and chemical effects not addressed in this paper could confirm or reject this hypothesis. From the observational perspective, ices in foreground clouds have been already presented in the literature indicating values close to the ice column density {\it plateau} estimated in this paper for H$_2$O and CO$_2$.

\end{enumerate}

\acknowledgments
The authors acknowledge the careful revision of the anonymous referee that improved the quality of this paper. WRMR thanks the S{\~{a}}o Paulo Research Foundation - FAPESP (grant\# PD2015/10492-3, BEPE2016/23054-7) for the financial support of this work and also FVE/UNIVAP. The research of WRMR also acknowledges support from the European Research Council (ERC) under the European Union's Horizon 2020 research and innovation programme through ERC Consolidator Grant ``S4F'' (grant agreement No~646908). SP thanks FAPESP (grant\# BEPQ2016/22018-7), FVE, CAPES and CNPq.

\bibliographystyle{yahapj}
\bibliography{References}

\appendix

\section{Deconvolution of the profile between 5.5$-$7.5 $\mu$\MakeLowercase{m}}
\begin{figure*}
\includegraphics[width=\textwidth,height=\textheight,keepaspectratio]{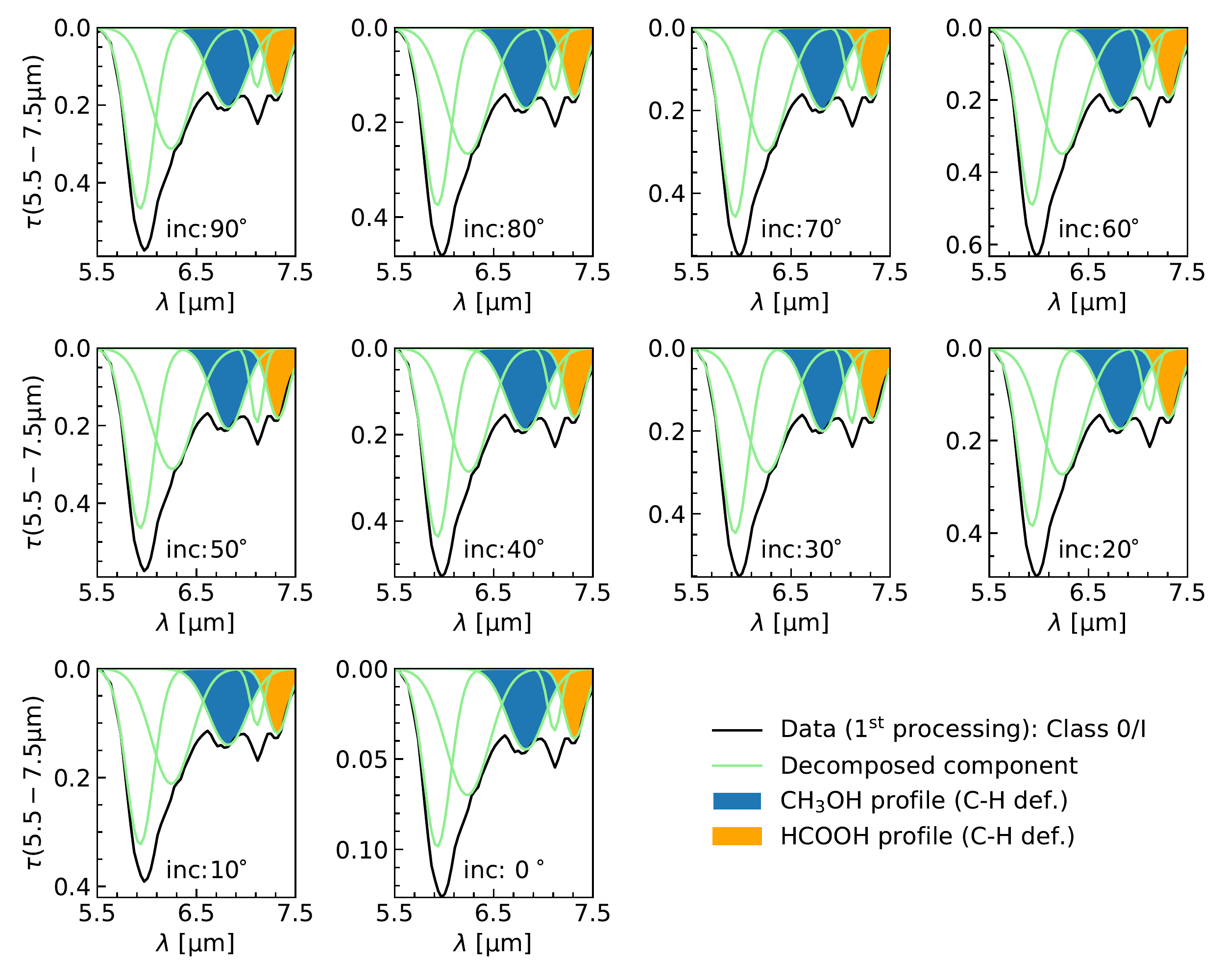}
\caption{Decomposition of the optical depth between 5.5$-$7.5~$\mu$m for the 1$^{st}$ processing model and Class 0/I. The black line shows the modelled data, whereas the green lines indicated the Gaussian profiles. The contribution of CH$_3$OH and HCOOH are given by the blue and orange areas, respectively.}
\label{fig:OpD_1_01}
\end{figure*}

\begin{figure*}
\includegraphics[width=\textwidth,height=\textheight,keepaspectratio]{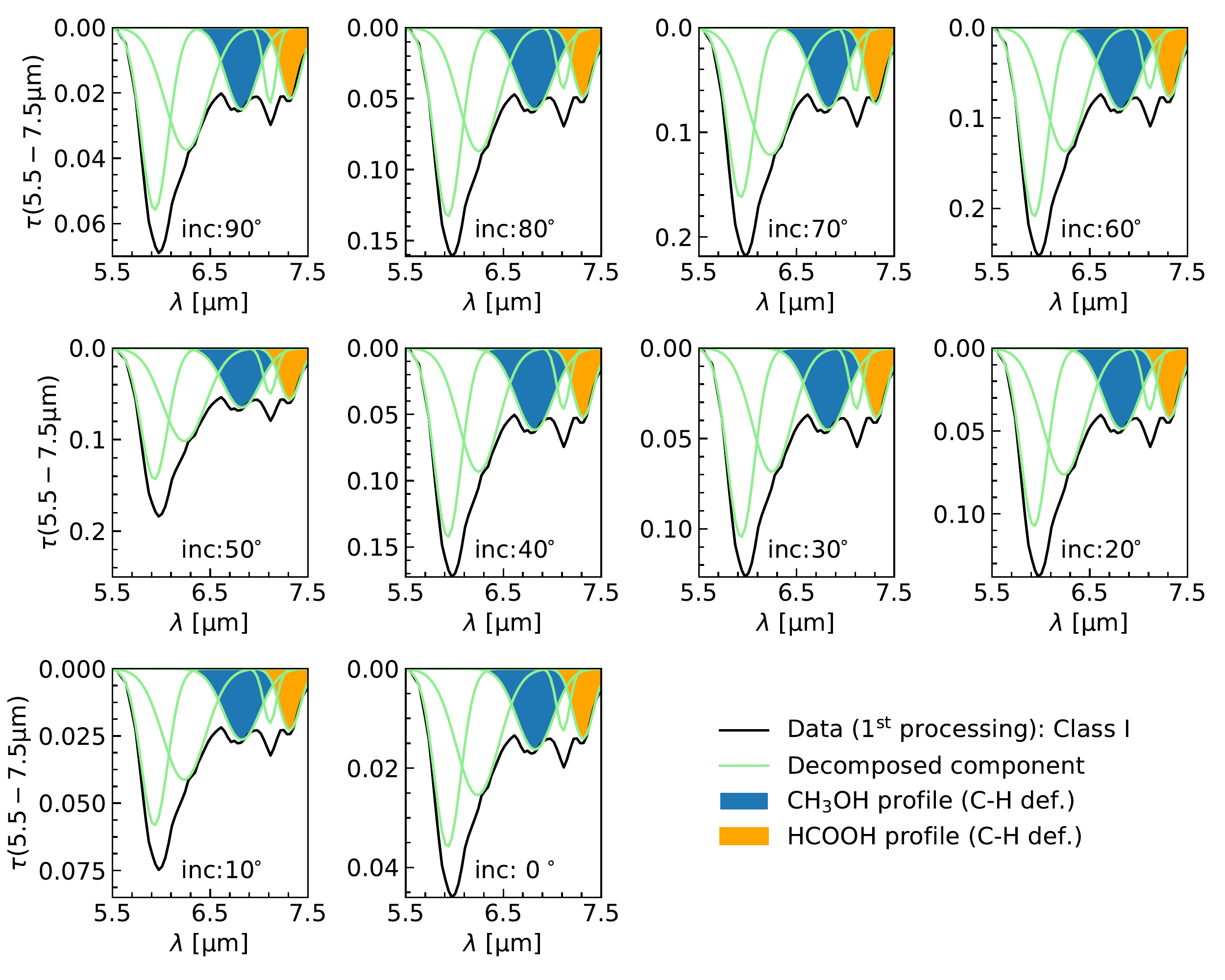}
\caption{Same of Figure 11 for Class I.}
\label{fig:OpD_1_1}
\end{figure*}

\begin{figure*}
\includegraphics[width=\textwidth,height=\textheight,keepaspectratio]{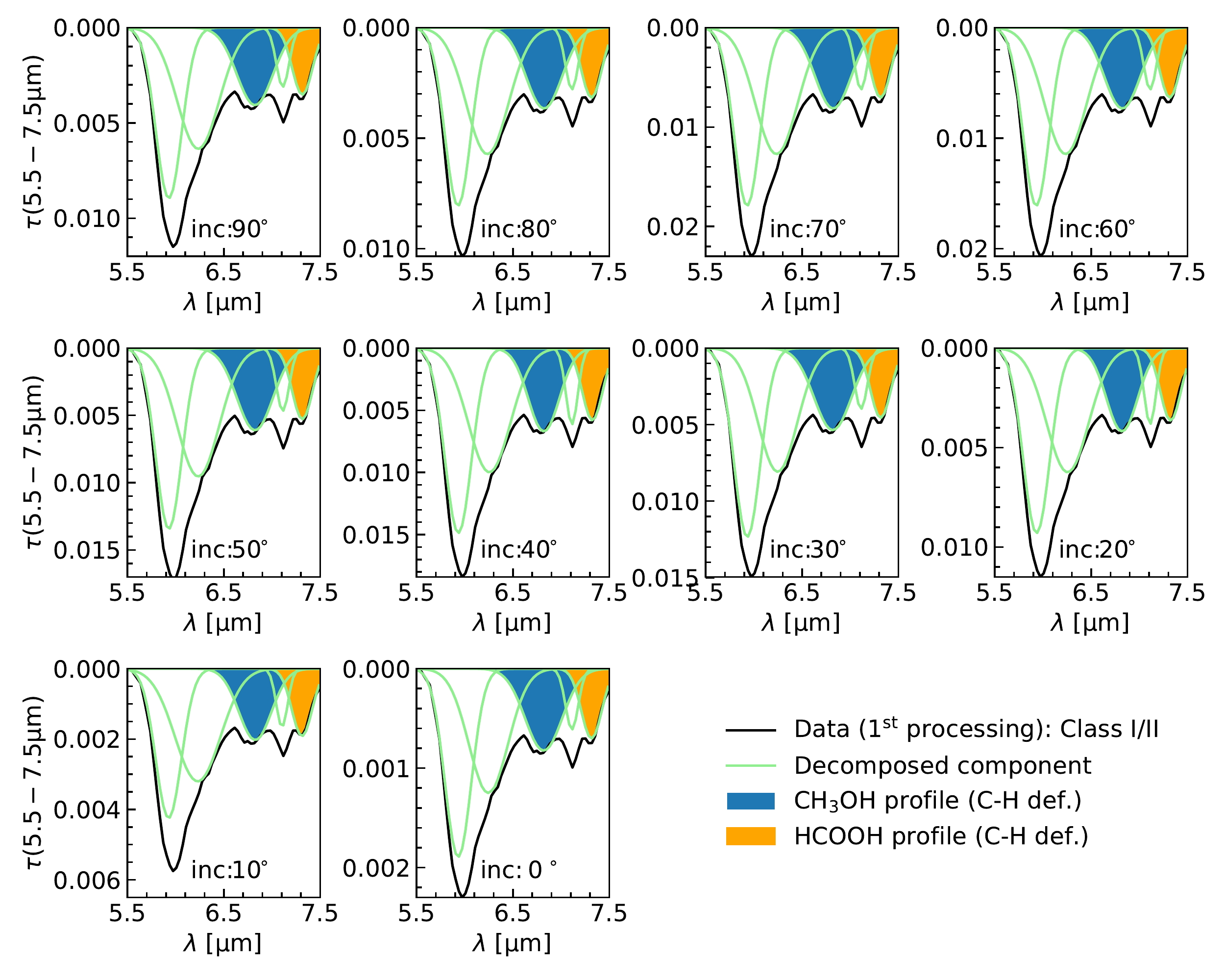}
\caption{Same of Figure 11 for Class I/II.}
\label{fig:OpD_1_12}
\end{figure*}

\begin{figure*}
\includegraphics[width=\textwidth,height=\textheight,keepaspectratio]{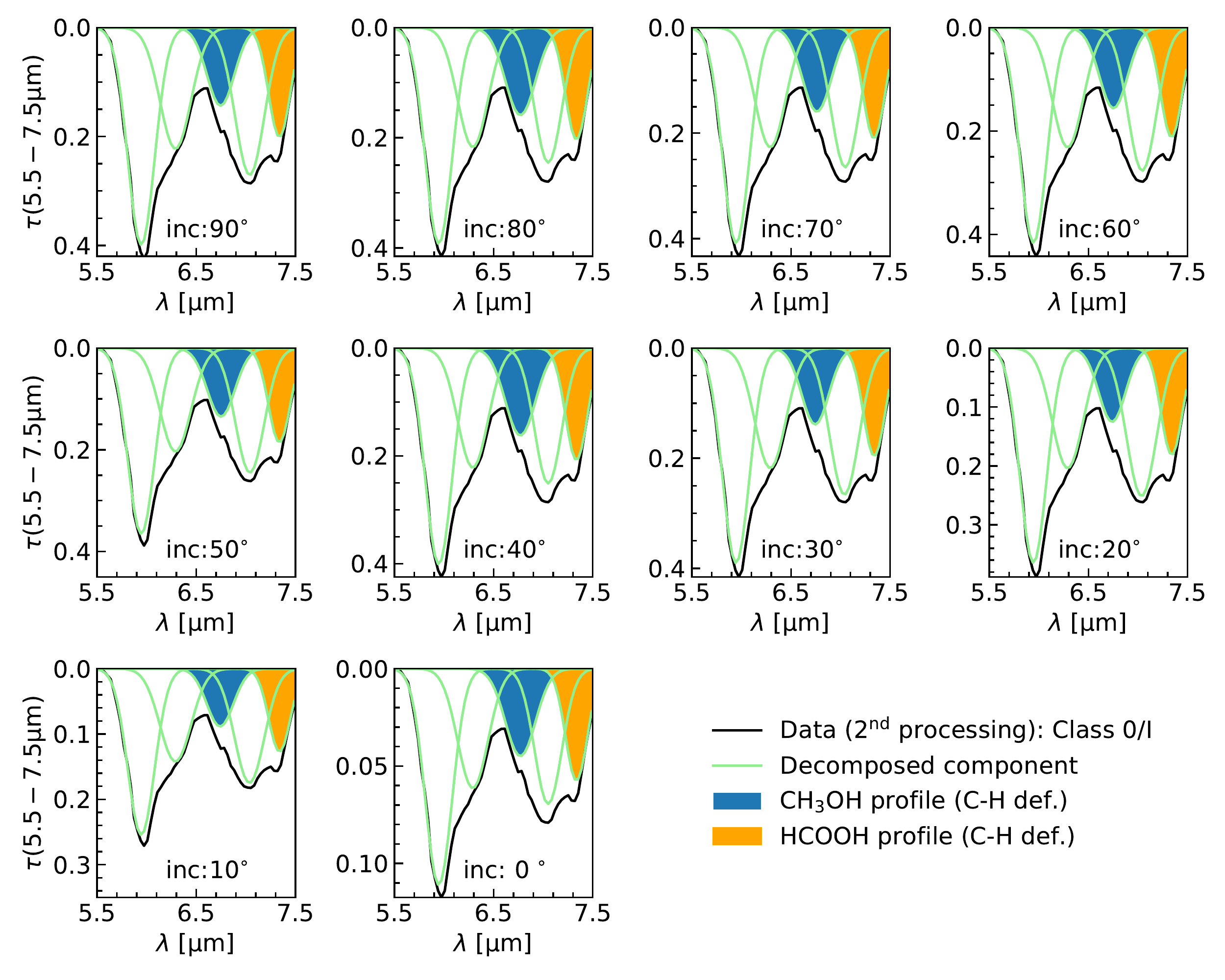}
\caption{Same of Figure 11 for the 2$^{nd}$ processing model and Class 0/I.}
\label{fig:OpD_2_01}
\end{figure*}

\begin{figure*}
\includegraphics[width=\textwidth,height=\textheight,keepaspectratio]{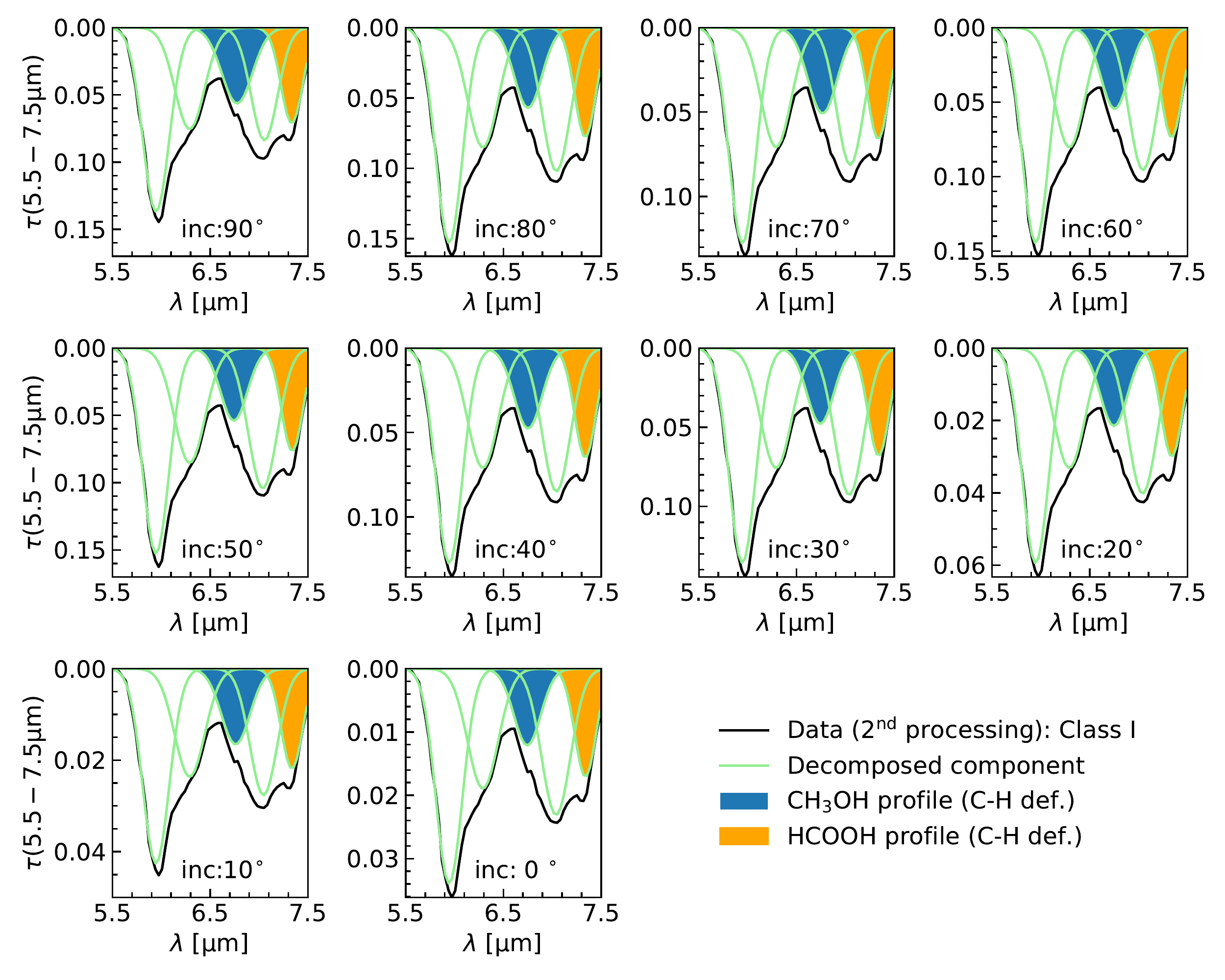}
\caption{Same of Figure 11 for the 2$^{nd}$ processing model and Class I.}
\label{fig:OpD_2_1}
\end{figure*}

\begin{figure*}
\includegraphics[width=\textwidth,height=\textheight,keepaspectratio]{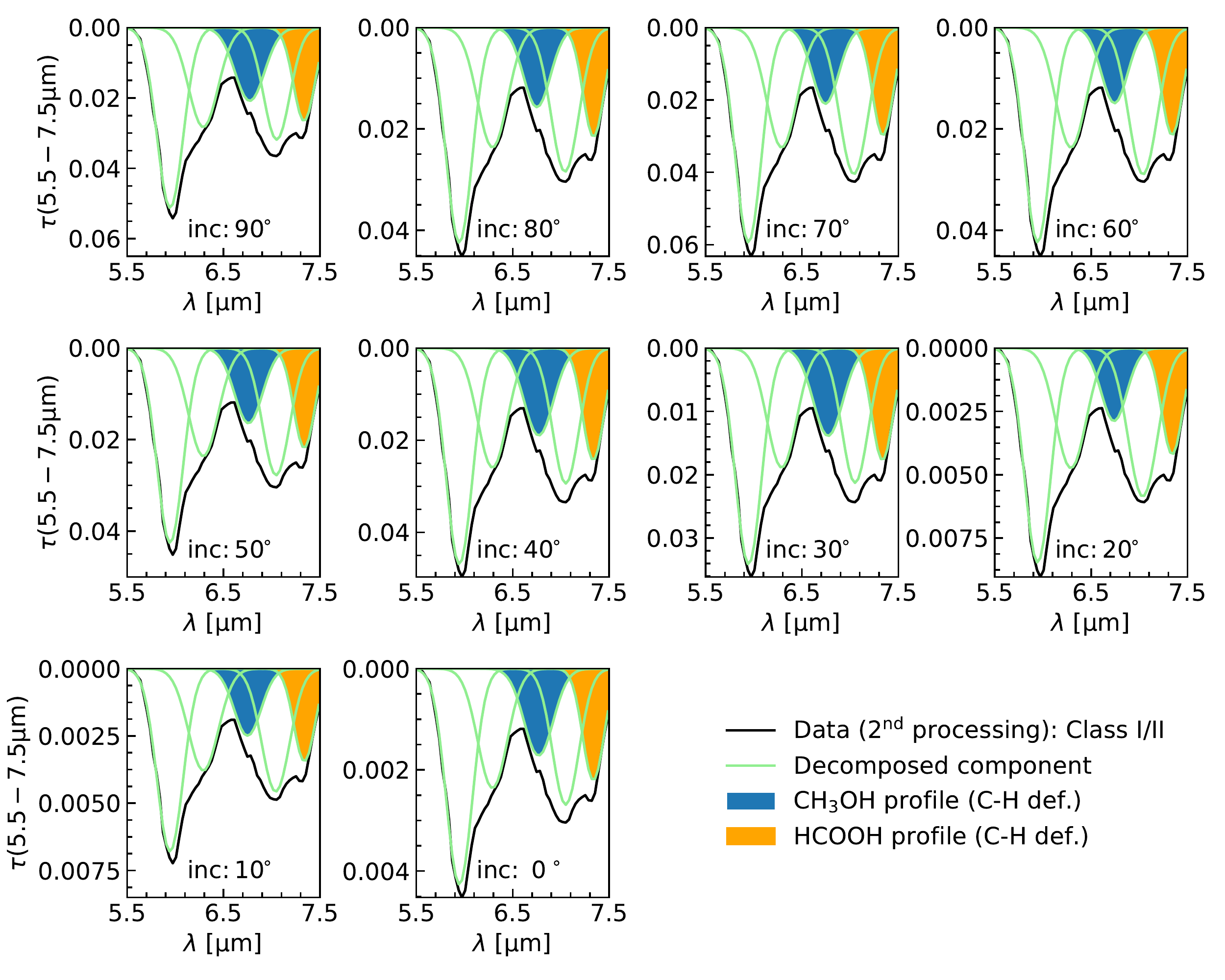}
\caption{Same of Figure 11 for the 2$^{nd}$ processing model and Class I/II.}
\label{fig:OpD_2_12}
\end{figure*}

The spectral region between 5.5$-$7.5 $\mu$m contains several O-H and C-H vibrational modes that might be associated to alcohols, ketone, and aliphatic ethers \citep{Schutte1996, Gibb2000, Keane2001, Boogert2008, Boogert2015}. In order to determine the contribution of CH$_3$OH and HCOOH in the spectral profile between 5.5$-$7.5~$\mu$m, a Gaussian decomposition of this spectral range by using 5 components was employed as given by:
\begin{subequations}
\begin{eqnarray}
  G(\lambda; A, \sigma) &=& \frac{A}{\sigma \sqrt{2\pi}} \mathrm{exp}\left[-\frac{(\lambda - \lambda_0)^2}{2\sigma^2}\right] \\
  \tau_{\lambda} &=& \sum_{\substack{i=0}}^{\substack{m-1}} G(\lambda^i; A^i, \sigma^i)
\end{eqnarray}
\end{subequations}
where $\lambda$ is the wavelength, $A$ is the integrated area, and $\sigma$ is the full width at half maximum of the Gaussian profile. Equation~A1a is the Gaussian function of one component and Equation~A1b is the sum of all components, assumed equal to 5 in this paper. Figures~\ref{fig:OpD_1_01}-\ref{fig:OpD_2_12} show the Gaussian decomposition of the spectral range between 5.5$-$7.5~$\mu$m according to the evolutionary stage variation and ice processing level. The two components around 6.0 $\mu$m are associated with H$_2$O and daughter species formed from the ice processing. The blue shaded component at 6.8~$\mu$m is attributed to CH$_3$OH whereas the yellow shaded component at 7.24~$\mu$m refers to HCOOH. The component in between ($\sim$7.0~$\mu$m) has been attributed to CH$_3$CHO in \citet{Pilling2010}.

\end{document}